\PassOptionsToPackage{pdfpagelabels=false}{hyperref} 
\documentclass[fleqn]{mnras}

\usepackage{natbib}
\usepackage{graphicx}	
\usepackage{amsmath}	
\usepackage{amssymb}	
\usepackage{color}
\usepackage{float}
\usepackage{mathptmx}
\usepackage{ulem}
\usepackage[euler]{textgreek}
\usepackage[utf8]{inputenc}
\usepackage{comment}
\usepackage{rotating}
\usepackage{booktabs}
\usepackage{tabularx}

\def\eq{\begin{equation}}
\def\en{\end{equation}}

\def\lesssim{\raisebox{-0.3ex}{\mbox{$\stackrel{<}{_\sim} \,$}}}
\def\gtrsim{\raisebox{-0.3ex}{\mbox{$\stackrel{>}{_\sim} \,$}}} 
\def\degsp{\hbox{$^{\circ}$}}
\def\ngj{n_{\rm GJ}}
\def\etal{{\it et al}\thinspace}
\def\cf{{\it cf.}\thinspace}
\def\apj{{\it Ap.J.}\thinspace}
\def\apjl{{\it Ap.J. Lett.}\thinspace}
\def\apjs{{\it Ap.J. Suppl.}\thinspace}
\def\aj{{\it A.J.}\thinspace}
\def\jaa{{\it J.Ap.A.}\thinspace}
\def\aap{{\it A\&A}\thinspace}
\def\aaps{{\it A\&A Suppl.}\thinspace}
\def\mnras{{\it MNRAS}\thinspace}
\def\P3hat{{\mathaccent 94 P}_3}
\def\parhang{\noindent\hangindent=0.4 true in \hangafter=1}
\def\nat{{\it Nature}\thinspace}
\def\deg{^{\circ}}
\def\etal{{\it et al.}\thinspace}
\def\ie{{\it i.e.,}\thinspace}
\def\eg{{\it e.g.,}\thinspace}
\usepackage{xcolor}

\title[Pulsar Beam Geometry of Broadband Arecibo Sources]{Pulsar Emission Beam Geometry of Radio Broadband Arecibo Sources\thanks{This paper is dedicated to our colleagues at the Institute for Astronomy, Kharkiv, Ukraine}}
\author[Olszanski, Rankin, Venkataraman, \&  Wahl]
{Timothy Olszanski$^{1, 2, 3}$\thanks{E-mail: timothyolszanski@gmail.com}, Joanna Rankin$^{3,4}$, Arun Venkataraman{$^5$}, Haley Wahl$^{1, 2, 3}$\\
$^{1}$Department of Physics and Astronomy, West Virginia University, P.O. Box 6315, Morgantown, WV 26505\\
$^{2}$Center for Gravitational Waves and Cosmology, West Virginia University, Morgantown, WV 26505\\
$^{3}$Physics Department, University of Vermont, Burlington, VT 05405, USA \\
$^{4}$Anton Pannekoek Institute for Astronomy, University of Amsterdam, Science Park 904, 1098 XH Amsterdam \\
$^{5}$Arecibo Observatory, bo. La Esperanza, P.P. Box 53995, Arecibo, Puerto Rico, 0612}

\date{Accepted XXX. Received XXX; in original form XXX}

\pubyear{XX}

\begin{document}
\label{firstpage}
\pagerange{\pageref{firstpage}--\pageref{lastpage}}
\maketitle

\begin{abstract}
We present radio pulsar emission beam analyses and models with the primary intent of examining pulsar beam geometry and physics over the broadest band of radio frequencies reasonably obtainable. We consider a set of well-studied pulsars that lie within the Arecibo sky. These pulsars stand out for the broad frequency range over which emission is detectable, and have been extensively observed at frequencies up to 4.5 GHz and down to below 100 MHz. We utilize published profiles to quantify a more complete picture of the frequency evolution of these pulsars using the core/double-cone emission beam model as our classification framework. For the low-frequency observations, we take into account measured scattering time-scales to infer intrinsic vs scatter broadening of the pulse profile. Lastly, we discuss the populational trends of the core/conal class profiles with respect to intrinsic parameters. We demonstrate that for this sub-population of pulsars, core and conal dominated profiles cluster together into two roughly segregated $P$-$\dot{P}$ populations, lending credence to the proposal that an evolution in the pair-formation geometries is responsible for core/conal emission and other emission effects such as nulling and mode-changing.

\end{abstract}




\section{Introduction}
Neutron stars that are detected through pulsed electromagnetic radiation are called pulsars. Radiation from pulsars is generated by plasma flowing along the magnetic field lines (flux tube) near the pulsar polar cap \citep{ruderman}. The physics and dynamics responsible for the plasma outflow and emission from pulsars are still an open and ongoing field of study \citep[\eg][]{Harding_2017,Melrose_2017,Melrose_2020,Cruz_2021}; thus, observational characteristics of the emission region are essential in constraining physical theories of the magnetosphere. Regarding radio emission mechanisms, many plausible suggestions have been offered, including streaming instabilities \citep{Melrose_2017,Ben_ek_2021}, orientation of the pair-formation-fronts (PFF) \citep{Cruz_2021}, and curvature radiation \cite{Harding_2017} to name a few. The difficulty remains in assessing which adequately captures all aspects of the radio emission behaviour we observe.  One important experimental constraint is the geometry of the radio emission region (hereafter referred to as the emission geometry).

Pulsars possess highly directional radiation beams, as the majority of the radiation generation is confined to the polar flux tube. An observer is limited to seeing intensity along a narrow line of sight crossing through this radiation pattern. The projections of the pulsar beam/beams onto an observer's line of sight are referred to as the main pulse and interpulse.  Integrating over a large number of rotations garners a stable measure of the pulsar's average radiation called the average profile\footnote{Average profiles are largely time-stable and characteristic of each individual pulsar, which make them useful for examining emission properties, pulse profile structure, and lastly the emission geometry.}, while substructure in the main pulse and interpulse are called components.   

The existence of components and their ordered frequency evolution amongst slower rotation-powered pulsars (hereafter canonical pulsars) provide the strongest evidence for an ordered emission geometry. Additionally, there is strong evidence indicating that the magnetic fields of canonical pulsars are approximately dipolar in the radio emission region (as seen in polarization properties such as \citet{Radhakrishnan}; it is very likely that the orderedness of the presenting emission geometry is intrinsically linked with the presence of field dipolarity. This assumption does not hold for other subpopulations, such as milli-second pulsars (MSPs)\footnote{Though there is evidence that a few MSPs have component structures similar to those seen in the canonical pulsar population and thus most likely possess approximately dipolar fields in the radio emission region. See \cite{Rankin_2017} for more details}, or magnetars.     
Efforts to quantify pulsar beams and their empirical properties remain a difficult task as we are limited to a confined (and randomized) sightline through each pulsar's beam configuration. Therefore, investigations have previously proceeded by proposing an underlying emission geometry and then attempting to identify scaling relations using a large ensemble of pulsars. The first beam model consisted of a single hollow cone \citep{Radhakrishnan, Komesaroff}.  Subsequent efforts added a central pencil beam and the possibility of two concentric hollow conical beams \citep{backer}, called the core/double-cone model. Early studies \citep{Cordes} of canonical pulsars suggested that the emission height was correlated with the emitted radiation's frequency (called radius to frequency mapping, hereafter RFM). Building on this work, \citet{rankin1983a} identified two different types of frequency evolution that components undergo, and attributed this effect to an ordered evolution of the emission geometry with emission height. 

Under this assumption, it is natural to infer that the frequency evolution of pulsar profiles is key to interpreting the underlying emission geometry.  However, the majority of pulsars have only been observed within a narrow portion of the radio band. Lower frequency observations are difficult both because of turnovers in pulsar spectra as well as the difficulty of correcting for dispersion smearing caused by the ionized Galactic interstellar medium (ISM) \citep{2004hpa..book.....L}.  In addition, broadening by scattering in the ISM often obfuscates component interpretation. Likewise, the majority of pulsars possess spectra that render them observationally undetectable at higher frequencies, thus limiting the frequency range over which frequency evolution can be detailed. Pulsar beaming studies then have historically focused on determining profile configurations around 1 GHz, including 
those conducted by \citet[hereafter ET VIa, ET VIb]{rankin1993a,rankin1993b}. 

A number of surveys now provide higher quality pulsar profiles that can then be used to study empirical trends over a larger frequency range and better constrain profile classifications. The Pushchino Radio Astronomy Observatory (PRAO) in Russia has long pioneered low-frequency studies of pulsar emission using their DKR-1000 cross telescope at frequencies below 100 MHz and the Large Phased Array (LPA) instrument at just above 100 MHz. Surveys by \citet[hereafter KL99]{kuzmin1999} and \citet[hereafter MM10]{malov} provide a foundation for this work.  Similarly, the 25/20-MHz survey by \citet[hereafter ZVK13+]{Zakharenko2013} using the Giant Ukrainian Radio Telescope (UTR-2) in Kharkiv follows a long record of pioneering pulsar observations in the decametric band.  More recently the Low Frequency Array (LOFAR) in the Netherlands has produced an abundance of high quality profiles starting with their High Band Survey \citep[hereafter BKK16+, PHS16+]{bilous2016, pilia} in the 100-200 MHz band and supplemented by others at lower frequencies \citep[hereafter BKK20++, BGT19+]{bilous2019, Bondonneau}. With respect to high frequency observations, the most recent example is the Arecibo 4.5-GHz polarimetric single-pulse survey by \citet[hereafter OMR19]{Olszanski2019} following older surveys such as \citet[hereafter vHKK97]{vonHoensbroech1997}.

Our overall goals of this paper and related papers \citep[\eg][]{paperiv} are to better understand the spectral changes in emission beam structure and to identify their physical implications. In this paper, we will primarily focus on assessing the efficacy of the core/double-cone beam model over the broadest frequency range yet. We focus on a large group of pulsars within the Arecibo sky that have been studied polarimetrically up to 4.5 GHz \citep[]{Olszanski2019} and observed down to 100 MHz or below.  These pulsars are well represented in the PRAO, UTR and LOFAR surveys, and most are seen in earlier surveys as \citet[hereafter GL98]{gould}, \citet[hereafter W99, W04]{Weisberg1999, Weisberg2004}, and \citet[hereafter HR10]{hankins2010}.  Almost all were discovered before 1975 and therefore have ``B'' names together with a long history of study which we both draw upon and reference as possible.  Readers can refer to OMR19 together with these works as needed to familiarize themselves with the basis for our analyses here.  

In what follows, \S 2 reviews the geometry and theory of core and conal beams, \S 3 describes how our beaming models are computed and displayed,  \S 4 discusses scattering and its effects at low frequency, \S 5 outlines a set of questions about core and conal emission for further consideration, and \S 6 summarizes the results. The main text details our analysis procedures while the tables, model plots and comments on individual sources are given in the Appendix. The supplementary material also includes ASCII-formatted files corresponding to the three tables.

\section{Core and Conal Beams}
\subsection{Pulse Profile Classification}
Canonical pulsar average profiles are observed to have up to five components \citep{rankin1983a}. This places an important constraint on the possible emission-beam configurations. Any beam model we adopt or postulate, must obey the condition that under a line of sight, no more than five components will ever be present in the main pulse. For this paper, we adopt the emission geometry proposed by \citet{backer}, which consists of two emission cones surrounding a central core emission beam; known as the core/double-cone model.

Under this model, pulse profiles divide into two major categories depending on which types of emission our line of sight encounters and thus which pulse components are most pronounced in the observed profile. For core profiles there are three types of possible pulse profiles: single (\textbf{S$_{t}$}) profiles consisting of an isolated core component\footnote{While this may seem to conflict with our adopted emission geometry, we would like to emphasize that core and conal components tend to have different spectral indices. Oftentimes, a pulsar classified as (\textbf{S$_{t}$}) grows conal components only at very high frequencies. It reiterates the point that broad frequency coverage is important in assessing the efficacy of postulated emission geometries.}, triple (\textbf{T}) profiles with a core component flanked by a pair of outriding conal components, or five-component (\textbf{M}) profiles with both an inner and outer pair of conal components and a central core component. By contrast, conal class profiles include conal single (\textbf{S$_{d}$}) which is comprised of a melded pair of cones, double profiles (\textbf{D}) consisting of a widely separated pair of conal components occasionally with a weak core component in-between,  or rarer intermediate geometries conal triple (c\textbf{T}) or quadruple (c\textbf{Q}) profiles. It should be noted that even if a profile is 
dominated by core or cone emission components, it doesn't preclude both from being present in the profile. Even so, the profile classification is important because it is a tracer of the prevalent type of emission encountered as our sightline traverses a pulsar's radiation beam. 

Each profile class tends to evolve with frequency in a characteristic manner:  core single (\textbf{S$_{t}$}) profiles often ``grow'' a pair of conal outriders at high frequency, whereas conal single (\textbf{S$_{d}$}) profiles tend to broaden and bifurcate at low frequency\footnote{
Both \textbf{S$_{d}$} and \textbf{S$_{t}$} can present very similar profiles, which underscores the importance of frequency evolution as a criterion in classification.}.  Triple (\textbf{T}) profiles usually show all three components over a very broad-band, but the relative component intensities can change greatly. Five-component (\textbf{M}) profiles tend to exhibit their individual components most clearly at meter wavelengths; at high frequency the components often become conflated and at low frequency the inner cone often weakens relative to the outer one.

Also important to the profile class is single-pulse phenomenology. Sub-pulse drift, a single-pulse phenomenon where emission appears to periodically ''drift'' across a phase window  has long been associated with conal emission and is another criteria in classifying a profile (as an example, see Fig.~\ref{figA109}). For a comprehensive review of sub-pulse drift and associated analyses, we refer the reader to \citet{2019MNRAS.482.3757B}. Highly periodic drifting, as well as other periodic emission can be folded to find an average modulation sequence. This is useful for studying highly periodic phenomena in our single-pulses and can be useful in identifying profile sub-structure that otherwise would be hidden in the average profile. A significant number of pulsars are well-modeled by the core/double-cone emission geometry, as described in a continuing series of publications with the overall title \textit{Empirical Theory of Pulsar Emission}, such as ET VI \citep{rankin1993a,rankin1993b}.

The two key angles describing the geometry are the magnetic colatitude $\alpha$ (angle between the rotation and magnetic axes) and the sightline-circle radius $\zeta$ (the angle between the rotation axis and the observer’s sightline), which relates to the sightline impact angle as  $\beta$ = $\zeta-\alpha$. Assuming a dipolar field and using spherical geometry, emission cone radii can be related to the conal component's half-power profile widths together with $\alpha$ and $\beta$ using 
\begin{equation} \label{eq1}
    \rho = \text{cos} ^{-1} [ \text{cos }\beta  - 2\text{sin }\alpha \text{ sin }\zeta \text{ sin }^{2} (\Delta\psi/4)]
\end{equation}
where $\Delta\psi$ is the total half-power width of the conal components measured in degrees \citep{rankin1993b}. The core beam is shown to scale directly with the core component's half-power profile width and is well modeled as a bivariate Gaussian (von Mises) beamform \citep{rankin1990}.

An important property of core and conal components are their empirically derived scaling relations. The three beams are found to have specific angular dimensions at 1 GHz in terms of the polar cap angular diameter, {$\Delta_{PC}$} = $2.45\degr P^{-1/2}$ \citep{rankin1990}.  The outside half-power radii of the inner and outer cones, {$\rho_{i}$} and {$\rho_{o}$}, are $4.33\degr P^{-1/2}$ and $5.75\degr P^{-1/2}$, and were determined experimentally by \citet{rankin1993a}.  Other studies such as \citet{gil}, \citet{kramer}, \citep{Bhattacharya}, and \citep{mitra1999} have come to very similar conclusions. For core components, $W_{\rm core}$, has empirically been shown to scale as 
\begin{equation}
W_{core}=\Delta_{\rm PC}/\sin\alpha = 2.45\degr P^{-1/2}/\sin\alpha
\end{equation}
as described in \citet{rankin1990}.

In practice, the magnetic colatitude $\alpha$ can be estimated using the aforementioned scaling relation if a core component is present in the $\sim$1-GHz profile.   The sightline impact angle $\beta$ can then in turn be estimated from the steepest gradient of the polarization position angle $\chi$ (PPA) traverse (at the inflection point in longitude $\varphi$), where $R$=$|d\chi/d\varphi|$ measures the ratio $\sin\alpha/\sin\beta$. The emission characteristic heights can then be computed assuming dipolarity using
\begin{equation}
r_{cone}\text{[km]} = r_{core}(\rho_{cone}/\rho_{core})^{2} = 6.66 \rho_{cone}^{2}P
\end{equation}
where $r_{core}$ is assumed to be 10 km \citep{rankin1993b,rankin1990}

These 1-GHz inner and outer conal emission heights have typically been seen to concentrate around 130 and 220 km, respectively.  However, it is important to recall that these are {\it characteristic} emission heights, not physical ones, estimated using the convenient but perhaps problematic assumption that the emission occurs adjacent to the ``last open'' field lines at the polar flux tube edge.  More physical emission heights can be estimated using aberration/retardation \citet{Blaskiewic} as corrected by \citet{dyks}), and these are typically somewhat larger than the characteristic emission heights.

A number of studies have followed expanding the population of pulsars with classifications. Most have looked at the  frequency evolution between $\mathtt{\sim}${300} MHz and $\mathtt{\sim}$1500 MHz in order to model a pulsar's emission geometry at 1 GHz \citep{Weisberg1999, Weisberg2004, mitra2011, brinkman_freire_rankin_stovall}. \citet{Olszanski2019} took a similar approach but extended consideration to 4.5 GHz.  Other compendia of polarized pulsar profiles, such as \citet{gould, han2009}, and more recently \citet{mcewen} have also been useful.  

\begin{figure}
\begin{center}
\includegraphics[width=90mm,angle=0.]{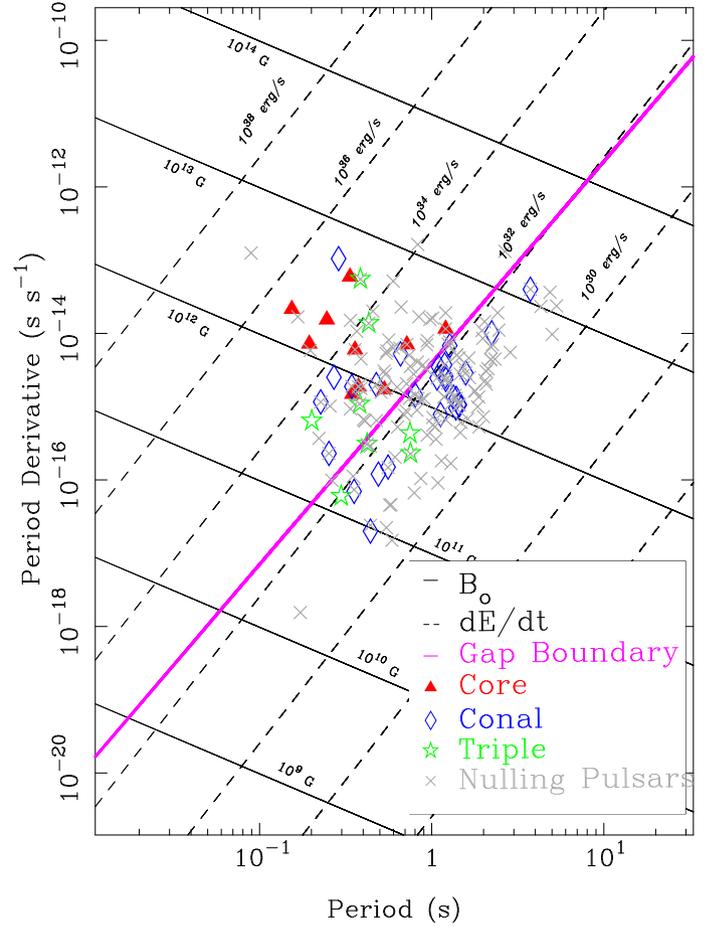}
\caption{$P$-$\dot P$ Diagram showing the location of core, conal, and triple profiles with respect to the PFF boundary line. Also plotted are pulsars known to null. Core pulsars tend to lie to the upper left of the boundary line while conal emission falls to the lower right (see text). Triple profiles with their mixture of core and conal emission, lie in between. Only one conal source falls significantly far away from the boundary line, J0631+1036. This source 
provides an example of an unusual c\textbf{Q} profile, and clearly stands out as abnormal compared to the majority of our sources. The great majority of nulling pulsars sit relatively close to the given boundary line.}
\label{fig1}
\end{center}
\end{figure}

\subsection{Generation of Profile Types by Plasma Source}
The outflowing electron-positron plasma that gives rise to a pulsar's emission is partly and or fully generated by a polar "gap" \citep{ruderman}, just above the stellar surface.  \citet{Timokhin} find that this plasma can be generated in one of two PFF configurations.  For the younger, energetic part of the pulsar population, pairs are created at some 100 m above the polar cap in a central, uniform (1-D) gap potential that produces copious backflow heating and thus thermal X-rays, forming a 2-D PFF; whereas for older pulsars the pair-formation front has a lower, annular shape and extends up along the conducting walls of the polar flux tube, thus becoming three-dimensional (cup shaped) with a 2-D gap potential and greatly reduced backflow heating. Curvature radiation generates the pair plasma in both cases, dominating the inverse-Compton process. An approximate boundary line between the flat and cup-shaped pair-formation geometries---and thus pulsar populations---is plotted on the $P$-$\dot P$ diagram of Fig.~\ref{fig1}, so that the more energetic pulsars are to the top left and those less so at the bottom right.  Its dependence is $\dot P=$4.29$\times10^{-29} \rho_{cm,7} P^{9/4}$, where $\rho_{cm,7}$ is the fieldline curvature in units of $10^{7}$ cm.  
The dependence for fieldline curvature is given as $\rho$ = 9.2$\times$10$^7 P^{1/2}$, which overall gives\footnote{Please note that there is a error in Eq. 59 of \citet{Timokhin}. The correct factor is $10^{-15}$, as we have given in this work.} 
\begin{equation}
\dot{P}= 3.95 \times 10^{-15}P^{11/4}
\end{equation}

This division is fundamental to the core/double-cone beam model of ET VI. Pulsars with conal single (\textbf{S$_d$}), double (\textbf{D}) and five-component (\textbf{M}) profiles tend to fall below the boundary line to the right, whereas those with core-single (\textbf{S$_t$}) profiles are found to 
the upper left of the boundary.  Those with triple (\textbf{T}) profiles are found on both sides of the boundary.  In the parlance of ET VI, the division corresponds to an acceleration potential parameter $B_{12}/P^2$ of about 2.5, which in turn represents an energy loss $\dot E$ of 10$^{32.5}$ ergs/s.  This delineation also squares well with \citet{Weltevrede2008}'s observation that high energy pulsars have distinct properties and \citet{basu2016}'s demonstration that conal drifting occurs only for pulsars with $\dot E$ less than about $10^{32}$ ergs/s.

It's important to point out that the boundary shown in Fig. \ref{fig1} is a theoretical result and individual pulsars are likely to depart somewhat depending on their own magnetospheric conditions. As a pulsar ages through this threshold, it's reasonable to expect that the emission dynamics would be prone to transitory mixed behaviours such as mode-changing, nulling, and amplitude modulation (all indicative of changes in plasma supply and generation). Mode-changing in particular alters the core/cone dominance and changes how a profile is interpreted.  Strong evidence of this has already been seen in B0823+26, which sits directly on the boundary line and has been one of the best studied core-single pulsars in the Arecibo sky. Only recently was the pulsar discovered to rarely exhibit a weak second mode\footnote{It's also worth noting that this pulsar undergoes amplitude modulation, periodic nulling, and a non-sudden transition between modes.}, which features a mainly conal profile  \citep{rankin_olszanski2020}\footnote{Partial profiles or single pulse plots such as the top half of the left panel in Fig. 1 of \citet{basu2018} show this more explicitly.} with traces of core emission, similar to a specific subpopulation of conal profiles \citet{young}. Had its secondary emission mode been more frequent, this pulsar might have been classified as conal. It's further worth noting that the great majority of pulsars known to exhibit nulling \citep{Konar_2019} concentrate around this region of the $P-\dot{P}$ diagram. \citet{10.1111/j.1365-2966.2007.11703.x} further studied the nulling fraction, and pulsars with higher nulling fractions appear to lie close to and around this boundary line as shown in their Fig. 2. If mode-changes are associated with a pulsar's transition through this boundary, it could explain the odd blending of core/conal/triple profiles we see in Fig. \ref{fig1}, and suggest some of these pulsars have yet to be discovered secondary emission modes.

\section{Computation and Presentation of Geometric Models}
Two observational values are key to computing conal radii: the profile width and the polarization position-angle sweep rate $R$. The former gives the angular scale of the emission beam, and latter the impact angle $\beta$ showing how the sightline crosses the beam.  Fig. 2 of ET VI depicts this configuration and the spherical geometry underlying the emission.

Empirically, core radiation beams are found to have bivariate Gaussian (von Mises) beamforms such that their invariant widths (FWHM) measure $\alpha$ but provides no $\beta$ information. If a pulsar has a core component, we attempt to use its width at around 1-GHz to estimate the magnetic colatitude $\alpha$, and when this is possible the $\alpha$ value is bolded in Table~\ref{tabA3}.  $\beta$ is then estimated from $\alpha$ and the polarization position angle sweep rate $R$ at the point of steepest gradient. The outside half-power (3 db) widths of conal components or pairs are measured, and the spherical geometry above then used to estimate the outside half-power conal beam radii.  Where $\alpha$ has been measured, that value is used, otherwise a value is estimated by using the established conal radius or characteristic emission height for an inner or outer cone at 1 GHz.  These conal radii and core widths are computed for different frequencies where possible.  

In most cases, our profile measurements follow closely from those in OMR19 and exhibit 1-GHz geometrical models identical or very similar to those given in their Table 6.  However, here we extend the analysis using LOFAR High Band 100-200-MHz or Pushchino Observatory 102/111-MHz profiles and in some cases below 100 MHz using LOFAR Low Band, Pushchino or UTR profiles. Table~\ref{tabA1} gives the sources for these profiles in both principal bands as well as each pulsar's observational parameters.  Table~\ref{tabA2} gives the physical parameters that can be computed from the period $P$ and spindown $\dot P$---that is, the spindown energy $\dot E$, spindown age $\tau$, surface magnetic field $B_{\rm surf}$, the acceleration parameter $B_{12}/P^2$ and the reciprocal of \citet{pulsar_magnetosphere_book}'s similar $Q$ (=$0.5\ 10^{15} \dot P^{0.4} P^{-1.1}$) parameter.  

Following the analysis procedures of ET VI, we have measured outside conal half-power (3 db) widths and half-power core widths where useful, using Gaussian fits from Michael Kramer's bfit code as detailed in \citet{1994A&AS..107..515K}.  However, we do not plot these directly.  Rather we use the widths to model the core and conal beam geometry in a manner similar to that of OMR19, but here emphasizing as low a frequency range as possible. The model results are given in Table~\ref{tabA3} for the 1-GHz band and for the respective 100-200 and $<$100-MHz band regimes.  

\begin{figure}
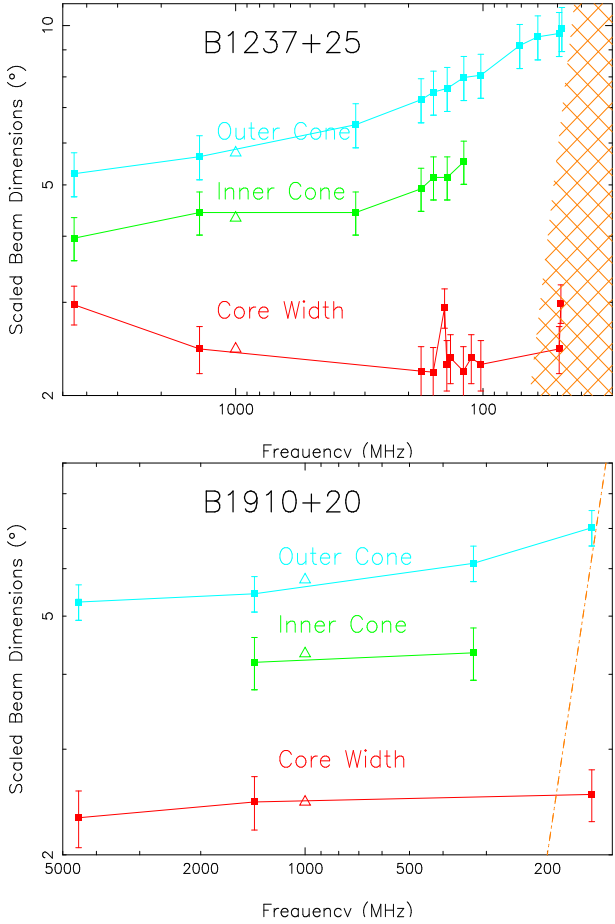

\begin{center}
\includegraphics[width=60mm,height=80mm,angle=-90.]{plots/B1237+25_Amodel.ps}
\includegraphics[width=60mm,height=80mm,angle=-90.]{plots/B1910+20_Amodel.ps}
\caption{Sample core/double-cone beam model displays for pulsars B1237+25 and B1910+20.  Curves for the scaled outer and inner conal radii and core width are shown---the former by $\sqrt{P}$ and the latter by $\sqrt{P}\sin\alpha$.  Conal error bars reflect the rms of 10\% uncertainties in both the profile widths and PPA rate (see text), and the core errors also reflect a 10\% error.  The upper display gives an example of a measured low frequency scattering level as indicated by the double hatching; whereas the lower display shows 10x the average scattering by the orange (dot-dashed) line. The plots are logarithmic on both axes.}
\label{fig20}
\end{center}
\end{figure}

We plot our results in terms of core and conal beam dimensions as a function of frequency, as described in Fig.~\ref{fig20}. The results of the model for each pulsar are then plotted in Figs.~\ref{figA1} to \ref{figA4}. 
For each pulsar the plotted values represent the \textbf{scaled} inner and outer conal beam radii and the core angular width, respectively.  
The conal beam radii are scaled by a factor of $\sqrt{P}$ and the core width (diameter) by $\sqrt{P}\sin{\alpha}$. This facilitates easy comparison of the beaming models for different objects as well as showing how each evolves with frequency relative to expected 1-GHz dimensions. The outer and inner conal radii are plotted with blue and green lines and the core diameter in red.  The nominal values of the three beam dimensions at 1 GHz are shown in each plot by a small triangle.  

Estimating and propagating the observational errors in the width values is a difficult task as errors can spawn from a variety of different sources. Examples include unknown instrumental errors, issues in interpreting and measuring the profile, inherent effects such as mode changing,  to name only a few. Therefore, we have chosen to show error bars reflecting the (scaled) beam-radii {$\rho_{\rm scaled}$} errors for a 10\% uncertainty in the width values and a 10\% uncertainty in the polarization position-angle (PPA) sweep rate---the former $0.1(1-\beta/\rho)\rho_{\rm scaled}$ and the latter $0.1(\beta/\rho)\rho_{\rm scaled}$. The error bars shown reflect the {\it rms} of the two sources with the former indicated in the lower bar and the latter in the upper one.  For many pulsars, only one of the errors is dominant so the bars corresponding to the two individual error sources are hard to see; however, B1633+24 provides a case where they can be seen quite clearly. The errors shown for the core-beam angular diameters are taken to be 10\% in the scaled width.

\section{Low Frequency Scattering Effects}
Scattering in the local interstellar medium distorts and broadens profiles by delaying a portion of the pulsar's signal.  This accrues as an exponential tail that can go from being hardly noticeable to dominant within an octave or so due to its steep ($f^{~-4}$) frequency dependence. Beam modeling relies upon the intrinsic profile dimensions, therefore it is important to include the effects of scattering in our analysis so as to better distinguish intrinsic frequency evolution from scatter broadening.

Many of the pulsars in this group have published scattering or scintillation studies that can be used to estimate the scattering time at a given frequency.  We are indebted to \citet{kuzmin2007,kuzmin_LL2007} for their extensive compendium of 100-MHz scattering times as well as other studies by \citet{alurkar}, \citet{geyer} and \citet{Zakharenko2013}. We attempt to estimate the regime where scattering significantly contributes to the growth of beam dimensions. In cases where 100 MHz scattering time-scales are available, they are converted to rotational phase and scaled by $\sqrt{P}$ to make the scattering comparable to core beam dimensions. The frequency evolution is then scaled using an index of $-4.1$ as adopted by \citet{kuzmin2007} and are shown on the model plots as double-hatched orange regions where the boundary reflects the scattering timescale as a function of frequency (\eg see the model plots in Fig.~\ref{figA1}). It is important to emphasize that the scattering timescale is being plotted alongside the scaled beam dimensions. As cone radii are related to their respective half-power widths through a geometric relationship, the way we have scaled the scattering will make it directly comparable to core widths, while serving only as an estimate for scattering broadening of the conal beam dimensions.\footnote{To estimate this properly would require propagating the effects of scattering on the half-power widths using the relation given in  Eq. \ref{eq1}.} It is also important to note that the scattering line of sight can evolve significantly during timescales of months/years between observations causing fluctuations around a given mean. It is thus likely that the experimentally measured scattering timescales we have plotted are not applicable to all of the observation epochs used.    

For pulsars where no scattering study is available, we adopt 
\begin{equation}
t_{\rm scatt} =e^{-6.46}DM^{2.194}f^{-3.86}
\end{equation}
 where frequency has units of GHz, and the scattering timescale is in ms, as described in \citet{2004ApJ...605..759B}. While this population's scattering level is well-determined, actual levels can depart from the average by up to an order of magnitude. Therefore, our model plots show the average scattering level (where applicable) as yellow single hatching and with an orange line indicating 10 times this value as a rough upper limit (\eg see the model plot for pulsar B1848+13 in Fig.~\ref{figA3}).

Because of the scattering timescales dependence, the expected core component beam dimension should show a steady growth towards lower frequencies once past the scattering boundary and equivalently, the conal beam dimensions will also grow\footnote{$\alpha$ is positive, and there are no cases for the estimated geometries where $\zeta$ will be negative}. Generally, the beam dimensions in the scattering regime behave as expected with our expectations for the majority of pulsars studied. Careful attention should be drawn to B0523+11, and B1633+24, where there is no apparent change in beam dimensions even in the scattering regime. We believe that for these cases, it is most likely that the scattering timescale decreased to a significantly lower value by the time low-frequency observations were taken.

Aside from a few outliers, the majority of inner cones show very little frequency evolution. The outliers can and most likely are explained by misclassification and scattering effects, however at least one appears to be intrinsic (B1237+25). Similarly, the vast majority of outer cones exhibit frequency evolution, while again there are several outlier cases where no frequency evolution is apparent. Core components appear to show no uniform trends.


\section{Exploring the Plasma Physics of Core and Conal Beams}
The core/double-cone model has been largely successful in providing a structure for identifying pulsars having similar beam configurations and in giving a basis for estimating a pulsar's magnetic colatitude $\alpha$ and sightline impact angle $\beta$ (\eg ET VI).  
Practical problems arise with conflated components and difficulties in estimating or interpreting the PPA rate. More significantly, many pulsars with rotational periods of 100 ms or less---and a few with longer periods---have complex profiles that are not well described by the core/double-cone model.  One possibility is that conformance to the model requires a dipolar magnetic field in the emission region---and this may well be the case for most slow pulsars at several hundred km emission heights, but not for some faster pulsars where multipole fields cause deformations to the underlying emission geometry.  

The larger question is what conformance to the core/double-cone model---seen polarimetrically and across the entire pulsar emission band---implies about the plasma physical processes that underlie the emission seen in these pulsars.  One or two sources of e$^-$/e$^+$ plasma produce the host of fascinating dynamical processes that have been identified over the last half century such as subpulse drifting, mode-changing, nulling \citep[\eg][ET III]{et3}, and these in turn generate the profile forms we observe with such effects as anti-symmetric $V$ signatures (\eg ET IV/V, \citep[\eg][ET IV/V]{rankin1990,etv}), edge depolarization \citep[][ET VIII]{et8} and 90\degr\ linear polarization (modal dominance) ``flips''.  These effects are manifested at the radio emission height within the polar flux tube, and apparently this height changes with wavelength as the flux tube opens and \textbf{B} decreases along with both the plasma density and plasma frequency.
 
Comprehensive analyses and interpretations of radio pulse profile variations with radio frequency are thus crucial. Pulsar beams together with their spectral and polarization changes carry uniquely important information about the geometry of the polar flux tube and plasma physical conditions within  that give rise to the radio emissions we receive.

\subsection{Core Beams}
Core components are observed to have a 1-GHz longitude width W$_{core}$=2.45\degr$P^{-1/2}\csc\alpha$ reflecting the 2.45\degr$P^{-1/2}$ angular diameter of the polar cap.  The 1-GHz core width has proven to be a highly reliable benchmark amid a sea of irregular pulsar parameters \citep{rankin1983a}; however, variations in core widths are observed that are not well-understood. For instance, the well-studied B1237+25, has two parts to it's core component with the trailing section usually much stronger than the leading, such that its width below in Fig.~\ref{figA2} remains nearly constant. On the other hand, another well-studied pulsar, B0823+26 (Fig.~\ref{figA2}) has a core width that seems to increase very substantially at lower frequencies.  Given the well-established connection between the core width and the polar cap diameter, any variation in the core width needs improved understanding.

Core components tend to have a distinct asymmetry between their leading and trailing parts.  
Even those with near Gaussian forms and prominent antisymmetric circular polarization ($V$) tend to have a discernible asymmetry in the polarization profile. In some, there is further evidence of an early and late orthogonal polarization mode (OPM). 
Further, some core components have have substantial linear polarization and PPA traverses that are smooth continuations of  adjacent conal components---but in other cases, the PPA appears disordered and linearly depolarized.  Isolated cores do not seem to show edge depolarization, and thus conal power may be responsible for any orderly PPA traverse. There is thus much to understand about the polarization of core components and the reader can observe examples of these circumstances as seen in \citep{Olszanski2019}.

Our low-frequency observations can also teach us about the relative RF spectra of the core and conal emission.  At high frequencies, core radiation often seems to have a softer spectrum than the conal emission (ET I) in the same pulsar, but it is unknown if that is also true at long wavelengths.  We see few clear cases of core emission at very low frequency, but then this is also true for conal radiation, so spectral turnovers must mitigate against both.  It will thus be important to pay close attention to pulsars where both survive to low frequencies.

\subsection{Inner and Outer Conal Beams}
Five-component pulsars tend to have steep PPA sweep rates and small $\beta/\rho$ which suggests a sightline traverse near the center of the beam pattern.  Such pulsars also permit $\alpha$ to be estimated from the core width, so the conal radii can be computed straightforwardly with spherical geometry as detailed previously. The detailed analyses in ET VI show that the outside half power radii of the inner and outer cones at 1 GHz for 1-s pulsars are 4.33 and 5.75\degr, respectively, and that these radii scale in the same manner as the core width---that is, as the inverse square root of the rotational periods ($P^{-1/2}$).  Remarkably, we further find that ($T$) pulsars with a single cone and core component---where we can also use the core width to estimate $\alpha$---clearly fall into two types: those with inner cones and those with outer cones.

A number of questions then devolve from these circumstances, namely why there are two (and only two) types of cones, and what are the geometric and/or physical circumstances that determine these specific conal dimensions. These dimensions point to specific emission heights along particular dipolar field lines within the polar flux tube (\eg the 130 and 220 km along the flux tube edge as mentioned above), but we don't yet fully understand just why these particular field lines and heights.  

One prominent difference between the cones is that inner cone radii tend to change little with frequency, whereas outer cones usually show a strong frequency dependence.  The reasons for this wavelength dependence, however, are not well understood.

A complication is that for many pulsars it is difficult to determine whether their cone is an inner or an outer one.  If our sightline misses the core beam (or it is absent), we only see a cone.  For a pulsar like B1133+16 (see Fig.~\ref{figA2}) that shows a large increase in cone size with wavelength, we tend to assume it is an outer cone; whereas for pulsars like B0834+06 or B1919+21 (Figs.~\ref{figA2},\ref{figA3}) their lack of growth over a very large band suggests inner cones.\footnote{However, this argument for B1919+21 is complicated by evidence that its emission involves a double cone \citep{rankin1993b}}  
A remarkable effect seemingly seen in both types of cones is edge depolarization:  it appears that the outer edges of many conal components are depolarized by equal amounts of modal polarization (ET VIII). We see this in both inner and outer cones when they occur on their own. However, in configurations where both the inner and outer cone are present, the depolarization is seen on the outer cone but not the inner.  The depolarization is often complete on the extreme outer edges, so this must involve a strong physical principle.  Related to this is the circumstance that many conal single profiles---where the sightline is tangential---are highly depolarized; whereas those involving a more central sightline traverse show more linear polarization

A final mystery is that the inner and outer cones do not seem to represent fully independent emission.  In configurations where subpulse drifting can be observed in both cones simultaneously (usually conal triple or quadruple profiles), the rotating ``beamlets'' seem to maintain a similar spacing and phase in each cone \citep[\eg][]{Hankins+Wolszczan}.

\section{Summary}
We begin a process of examining how pulsar beams appear at lower frequencies and interpreting these changes in terms of the  pulsar emission geometry. Explicitly, we assume the emission geometry for canonical pulsars is comprised of a core/double-cone beam.  Here we consider a group of well-studied pulsars within the Arecibo sky.  This group also includes most of the Arecibo pulsars that have been detected at frequencies below 100 MHz.  We take the opportunity not only to review the various beam models for these pulsars, but also to better constrain profile classifications using the broadband observations we have at our disposal, as well as to point out situations where the core/double-cone model is unsatisfactory. 
Lastly, we also demonstrate that for this sub-population of pulsars, core and conal dominated profiles cluster together into two roughly segregated $P$-$\dot{P}$ populations, lending credence to the proposal that an evolution in the pair-formation geometries is responsible for core/conal emission. Future work with broader sub-populations will confirm if this trend holds true to the overall canonical pulsar population. Several adjoining papers are in process, the first studying a large population of pulsars outside the Arecibo sky with low frequency detections \citet{paperiv}). Another two will include groups of less studied pulsars within the Arecibo sky \citep{paperiii,paperii}. 

\section{Observational Data availability}
The profiles will be available on the European Pulsar Network download site, and the pulses sequences can be obtained by corresponding with the lead author.

\section*{Acknowledgements}
 HMW gratefully acknowledges a Sikora Summer Research Fellowship. Much of the work was made possible by support from the US National Science Foundation grant 09-68296 and 18-14397. The Arecibo Observatory is operated by the University of Central Florida under a cooperative agreement with the US National Science Foundation, and in alliance with Yang Enterprises and the Ana G. M\'endez-Universidad Metropolitana. We especially thank our colleagues who maintain the ATNF Pulsar Catalog and the European Pulsar Network Database as this work drew heavily on them both.  This work made use of the NASA ADS astronomical data system.



%

\bibliography{bib.bib}

\newpage

\appendix
\setcounter{figure}{0}
\renewcommand{\thefigure}{A\arabic{figure}}
\renewcommand{\thetable}{A\arabic{table}}
\setcounter{table}{0}
\renewcommand{\thefootnote}{A\arabic{footnote}}
\setcounter{footnote}{0}

\section{Pulsar Tables, Models, and Classification Notes}

\begin{table*}
\caption{Observation Information. Given are basic pulsar parameters, and references for surveys that included observations near and below 100 MHz.}
\begin{tabular}{lccc|l|l}
\hline
 Pulsar & P & DM & RM  & References &  References \\
    (B1950) & (s) & ($pc/cm^{3}$) & ($rad$-$m^{2}$) & OMR19, GL98 plus ...  & \\
    \hline
    & & & &  \mbox{\textbf{($\geq$100 MHz)}} &   \mbox{\textbf{($\leq$100 MHz)}}\\
    \hline\hline
B0301+19	&	1.39	&	15.7	&	--8.3	& JKMG; KKWJ; KL99; BKK16; KLL07	& BKK19; KZU22	 \\
B0523+11	&	0.35    & 	79.3	&	+37		& KL99; JKMG; BKK16; KLL07	&	\\
B0525+21	&	3.75	&	50.9	&	--39.6	& KL99; BKK16; KLL07	& PHS16; BGT20	 \\
J0538+2817  &   0.14    &   39.57   &   +39     &      &       \\
B0540+23    &   0.25    &	77.7    &   +8.7 	& BKK16; KLL07	&	 \\
\\
B0609+37	&	0.30    & 	27.1	&	+23		& KKWJ; KL99; BKK16; KLL07	& BKK19	\\
B0611+22	&	0.33    &	96.9    &	+69		& JKMG; KKWJ; KL99; KLL07	&	\\
B0626+24	&	0.48	&	84.2	&	+69.5	& KKWJ; KL99; BKK16; KLL07	&	 \\
J0627+0649  &   0.35    &   86.6    &   +179    &      &       \\
J0631+1046  &   0.29    &   125.4   &   +137    & ZCW96; MM10; KLL07  &   \\
\\
B0656+14	&	0.38	&	13.9	&	+23.0	& KL99; BKK16; KLL07	&	 \\
B0751+32	&	1.44	&	40.0	&	--7.0	& KKWJ; BKK16; KLL07		& BKK19	 \\
B0823+26	&	0.53	&	19.5	&	+5.4	& KL99; HR10	& HR10: BKK19; PHS16; BGT19; KLL07; ZVK+	\\
B0834+06	&	1.27	&	12.9	&	+25.3	& JKMG; KKWJ; NSK15; KL99; BRG9; XBT+ & PHS16; BGT19; ZVK+	 \\
B0919+06	&	0.43	&	27.3	&	+29.2	& JKMG; KL99; HR10; PHS16; KLL07; XBT+ & HR10; BGT19; ZVK+	 \\
\\
B0943+10    &   1.10    &   15.4    &  	+13.3   & KL99; HR10; BKK16; KLL07 & HR10; BKK19; BGT19; PHS16; ZVK+        \\
B0950+08	&	0.25	&	3.0	&	--0.7		& KL99; HR10; KLL070; XBT+ & HR10; PHS16; BGT19; ZVK+	 \\
B1133+16	&	1.19	&	4.8	&	+4.0		& KL99; BKK16; KLL07; XBT+ & HR10: BKK19; PHS16; BGT19; ZVK+	 \\
B1237+25	&	1.38	&	9.3	&	--0.1		& KL99; HR10; BKK16; PHS16; COR86 & BGT19; BKK19; ZVK+	 \\
B1530+27	&	1.12	&	14.7	&	+1.0	& KL99; BKK16; PHS16; KLL07	 & BGT19; BKK19; ZVK+	 \\
\\
B1541+09	&	0.75	&	35.0	&	+21.0	& KKWJ; KL99; HR10; BKK16; KLL07	 & PHS16; BGT19; BKK19	 \\
B1604--00	&	0.42    &	10.7 &	+6.5		&KL99; KKWJ; PHS16; COR86	& PHS16; BGT19; ZVK+	\\
B1612+07	&	1.21	&	21.4	&	+40.0	& KKWJ; KL99; PHS16; KLL07	& BGT19; ZVK+	 \\
B1633+24	&	0.49	&	24.3	&	+31.0	& KL99; BKK16; KLL07	& PHS16; BGT19; BKK19; ZVK+	 \\
B1737+13	&	0.80	&	48.7	&	+64.4	& BKK16; JKMG; KKWJ; KL99; KLL07	& BGT19; BKK19	 \\
\\
J1740+1000  &   0.15    &   23.9    &   +23.8   & MM10 &   \\
B1821+05	&	0.75    &	66.8 &	+145		& KKWJ; KL99; HR10; PHS16	&	\\
B1839+09	&	0.38	&	49.2	&	+53.0	& KKWJ; BKK16; PHS16; COR86		& BKK19	 \\
B1842+14	&	0.38	&	41.5	&	+109.0	& JKMG; HR10; BKK16; COR86	& BGT19; BKK19	 \\
B1845--01   &   0.66    &   159.5   &   +580    & MM10; KMN+15 &   \\
\\
B1848+12	&	1.21	&	139  &	????		& BKK16; PHS16		&	\\
B1848+13	&	0.35	&	60.2 &	+146		& JKMG; PHS16; BKK16		&	\\
B1910+20	&	2.23	&	88.3	&	+148	& BKK16		&	\\
B1914+09    &   0.27    &   61.0    &   97.0    & KKWJ; PHS16; KLL07 &  \\
B1915+13    &   0.19    &   94.5    &   233.0   & KKWJ; PHS16; KL99; XBT+; KLL07 &  \\
\\
B1919+21	&	1.34	&	12.4	&	--16.99	& KL99; JKMG; BKK16; XBT+; KLL07 & HR10; PHS16; BGT19; BKK19; ZVK+ \\
B1923+04    &   1.07    &   102.2   & -39.5     & KKWJ; PHS16; KMN+15 &  \\
B1929+10	&	0.23	&	3.2     &	--6.9   & KL99; BKK16; XBT+; COR86	& HR10: BGT19; BKK19; PHS16; ZVK+	 \\
B1933+16    &   0.36    &   158.5   & --10.2    & HR10; XBT+; KLL07 & \\
B1935+25    &   0.20    &   53.2    &   +26     & MM10; KLL07    &   \\
\\
B1944+17	&	0.44	&	16.2 	&	--28. 	& PHS16; BKK16; COR86		&  ZVK+	 \\
B1946+35    &   0.72    &  129.1    &   +116    & BKK16; MM10; KMN+15 & \\
B1952+29    &   0.43    &   7.93    &   -18     & MM10; KMN+15    &   ZVK+ \\
B2016+28	&	0.56	&	14.2	&	--34.6	& KKWJ; KL99; HR10; XBT+; BKK16	 & BGT19; BKK19; ZVK+	 \\
B2020+28	&	0.34	&	24.6	&	--74.7	& KL99; HR10; BKK16; XBT+; KLL07	 & BGT19+; BKK19	  \\
\\
B2110+27    &   1.20    &   25.1   &   --37    	& KKWJ; KL99; BKK16; KLL07       &  BGT19;  ZVK+        \\
B2303+30    &   1.58    &   49.5    &  --75.5   &HR10; KKWJ; BKK16; PHS16; KLL07       & BGT19; BKK19; PHS16  \\
B2315+21    &   1.44    &   20.9    &  --37     &HR10 KKWJ; BKK16; PHS16; KLL07       &    ZVK+         \\
\hline
\end{tabular}
\label{tabA1}
\vskip 0.1in
Notes: BGT19: \citet{Bondonneau}; BKK16: \citet{bilous2016}; BKK19: \citet{bilous2019}; BRG99: \citet{brg99}; COR86: \citet{Cordes}; GL98: \citet{gould}; JKMG: \citet{JKMG08}; HR10: \citet{hankins2010}; KKWJ: \citet{Kijak1998}; KL99: \citet{kuzmin1999}; KLL07: \citet{kuzmin_LL2007}; KMN+15: \citet{kmn+15}; LDK+13: \citet{ldk+13}; KZU22: \citet{Kravtsov22}; MM10: \citet{malov}; NSK15: \citet{Noutsos}; OMR19: \citet{Olszanski2019}; PHS16: \citet{pilia}; XBT+: \citet{Xue+17}; ZVK+: \citet{Zakharenko2013}; ZCW96: \citet{Zepka}.  Values from the ATNF Pulsar Catalog \citep{ATNF}.
\end{table*}

\begin{table}
\setlength{\tabcolsep}{4pt}
\caption{Pulsar standard measured and calculated parameters.}
\begin{center}
\begin{tabular}{lccccccc}
\hline
 Pulsar &  P & $\dot{P}$ & $\dot{E}$ & $\tau$ & $B_{surf}$ & $B_{12}/P^2$ & 1/Q  \\
 (B1950) & (s) & ($10^{-15}$  & ($10^{32}$  & (Myr) & ($10^{12}$ &   &   \\
 & & s/s) & ergs/s) & & G) &   &    \\
\hline
\hline
B0301+19 & 1.3876 & 1.30 & 0.19 & 17.0 & 1.4 & 0.7 & 0.4 \\
B0523+11 & 0.3544 & 0.07 & 0.65 & 76.3 & 0.2 & 1.3 & 0.6 \\
B0525+21 & 3.7455 & 40.05 & 0.30 & 1.5 & 12.4 & 0.9 & 0.5 \\
J0538+2817&0.1432 & 3.67 & 490 & 0.6 & 0.7 & 35.8 & 7.1 \\
B0540+23 & 0.2460 & 15.42 & 409 & 0.3 & 2.0 & 32.6 & 7.0 \\
\\
B0609+37 & 0.2980 & 0.06 & 0.89 & 79.4 & 0.1 & 1.5 & 0.6 \\
B0611+22 & 0.3350 & 59.45 & 620 & 0.1 & 4.5 & 40.3 & 8.5 \\
B0626+24 & 0.4766 & 2.00 & 7.28 & 3.8 & 1.0 & 4.3 & 1.5 \\
J0627+0649&0.3466 & 1.70 & 16.0 & 3.2 & 0.8 & 6.5 & 2.0 \\
J0631+1036 & 0.2878 & 104.67 & 1700 & 0.0 & 5.6 & 67.0 & 12.6 \\
\\ 
B0656+14 & 0.3849 & 55.00 & 381 & 0.1 & 4.7 & 31.4 & 7.1 \\
B0751+32 & 1.4423 & 1.08 & 0.14 & 21.2 & 1.3 & 0.6 & 0.3 \\
B0823+26 & 0.5307 & 1.71 & 4.52 & 4.9 & 1.0 & 3.4 & 1.2 \\
B0834+06 & 1.2738 & 6.80 & 1.30 & 3.0 & 3.0 & 1.8 & 0.8 \\
B0919+06 & 0.4306 & 13.73 & 68.0 & 0.5 & 2.5 & 13.3 & 3.6 \\
\\ 
B0943+10 & 1.0977 & 3.49 & 1.04 & 5.0 & 2.0 & 1.6 & 0.7 \\
B0950+08 & 0.2531 & 0.23 & 5.60 & 17.5 & 0.2 & 3.8 & 1.3 \\
B1133+16 & 1.1879 & 3.73 & 0.88 & 5.0 & 2.1 & 1.5 & 0.7 \\
B1237+25 & 1.3824 & 0.96 & 0.14 & 22.8 & 1.2 & 0.6 & 0.3 \\
B1530+27 & 1.1248 & 0.78 & 0.22 & 22.9 & 0.9 & 0.7 & 0.4 \\
\\ 
B1541+09 & 0.7484 & 0.43 & 0.41 & 27.4 & 0.6 & 1.0 & 0.5 \\
B1604-00 & 0.4218 & 0.31 & 1.60 & 21.8 & 0.4 & 2.0 & 0.8 \\
B1612+07 & 1.2068 & 2.36 & 0.53 & 8.1 & 1.7 & 1.2 & 0.6 \\
B1633+24 & 0.4905 & 0.12 & 0.40 & 65.1 & 0.2 & 1.0 & 0.5 \\
B1737+13 & 0.8031 & 1.45 & 1.11 & 8.8 & 1.1 & 1.7 & 0.7 \\
\\ 
J1740+1000 & 0.1541 & 21.47 & 2316 & 0.1 & 1.8 & 77.5 & 13.3 \\
B1821+05 & 0.7529 & 0.23 & 0.21 & 52.6 & 0.4 & 0.7 & 0.4 \\
B1839+09 & 0.3813 & 1.09 & 7.76 & 5.5 & 0.7 & 4.5 & 1.5 \\
B1842+14 & 0.3755 & 1.87 & 14.0 & 3.2 & 0.8 & 6.0 & 1.9 \\
B1845-01 & 0.6594 & 5.25 & 7.20 & 2.0 & 1.9 & 4.3 & 1.5 \\
\\ 
B1848+12 & 1.2053 & 11.52 & 2.60 & 1.7 & 3.8 & 2.6 & 1.1 \\
B1848+13 & 0.3456 & 1.49 & 14.3 & 3.7 & 0.7 & 6.1 & 1.9 \\
B1910+20 & 2.2330 & 10.18 & 0.36 & 3.5 & 4.8 & 1.0 & 0.5 \\
B1914+09 & 0.2703 & 2.52 & 50.0 & 1.7 & 0.8 & 11.4 & 3.1 \\
B1915+13 & 0.1946 & 7.20 & 390 & 0.4 & 1.2 & 31.7 & 6.7 \\
\\ 
B1919+21 & 1.3373 & 1.35 & 0.22 & 15.7 & 1.4 & 0.8 & 0.4 \\
B1923+04 & 1.0741 & 2.46 & 0.78 & 6.9 & 1.6 & 1.4 & 0.7 \\
B1929+10 & 0.2265 & 1.16 & 39.3 & 3.1 & 0.5 & 10.1 & 2.7 \\
B1933+16 & 0.3587 & 6.00 & 51.0 & 0.9 & 1.5 & 11.5 & 3.2 \\
B1935+25 & 0.2010 & 0.64 & 31.0 & 5.0 & 0.4 & 9.0 & 2.4 \\
\\ 
B1944+17 & 0.4406 & 0.02 & 0.11 & 290 & 0.1 & 0.5 & 0.3 \\
B1946+35 & 0.7173 & 7.06 & 7.55 & 1.6 & 2.3 & 4.4 & 1.6 \\
B1952+29 & 0.4267 & 0.00 & 0.01 & 3950 & 0.0 & 0.1 & 0.1 \\
B2016+28 & 0.5580 & 0.15 & 0.34 & 59.7 & 0.3 & 0.9 & 0.4 \\
B2020+28 & 0.3434 & 1.89 & 18.5 & 2.9 & 0.8 & 6.9 & 2.1 \\
\\ 
B2110+27 & 1.2029 & 2.62 & 0.59 & 7.3 & 1.8 & 1.2 & 0.6 \\
B2303+30 & 1.5759 & 2.89 & 0.29 & 8.6 & 2.2 & 0.9 & 0.5 \\
B2315+21 & 1.4447 & 1.05 & 0.14 & 21.9 & 1.2 & 0.6 & 0.3 \\

\hline  
\end{tabular}
\end{center}
\label{tabA2}
\vskip 0.1in
Notes: Values from the ATNF Pulsar Catalog \citep{ATNF}.
\end{table}

\begin{figure*}
\begin{center}
\includegraphics[width=175mm,height=222mm,angle=0.]{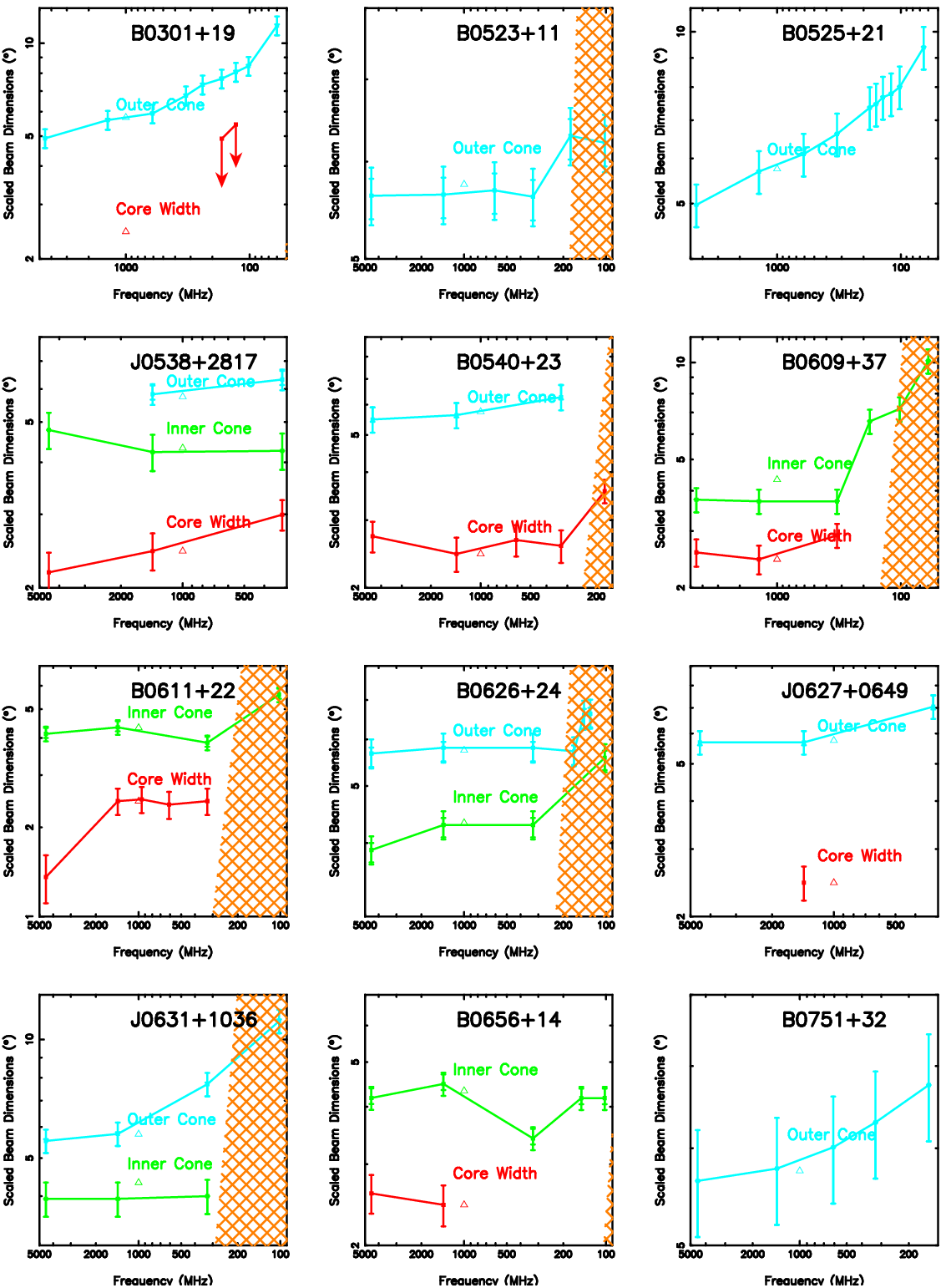}
\caption{Core/double-cone emission-beam model displays for pulsar B0301+19 through B0834+06.  The \textbf{scaled} angular dimensions of the outer cone (blue), inner cone (green) and core (red) beams are plotted together with errors reflecting 10\% uncertainties in the relevant measurements (see text).  Scattering levels are shown in orange double hashing when measurements are available and yellow single hashing with a 10X boundary when not.}
\label{figA1}
\end{center}
\end{figure*}

\begin{figure*}
\begin{center}
\includegraphics[width=175mm,height=222mm,angle=0.]{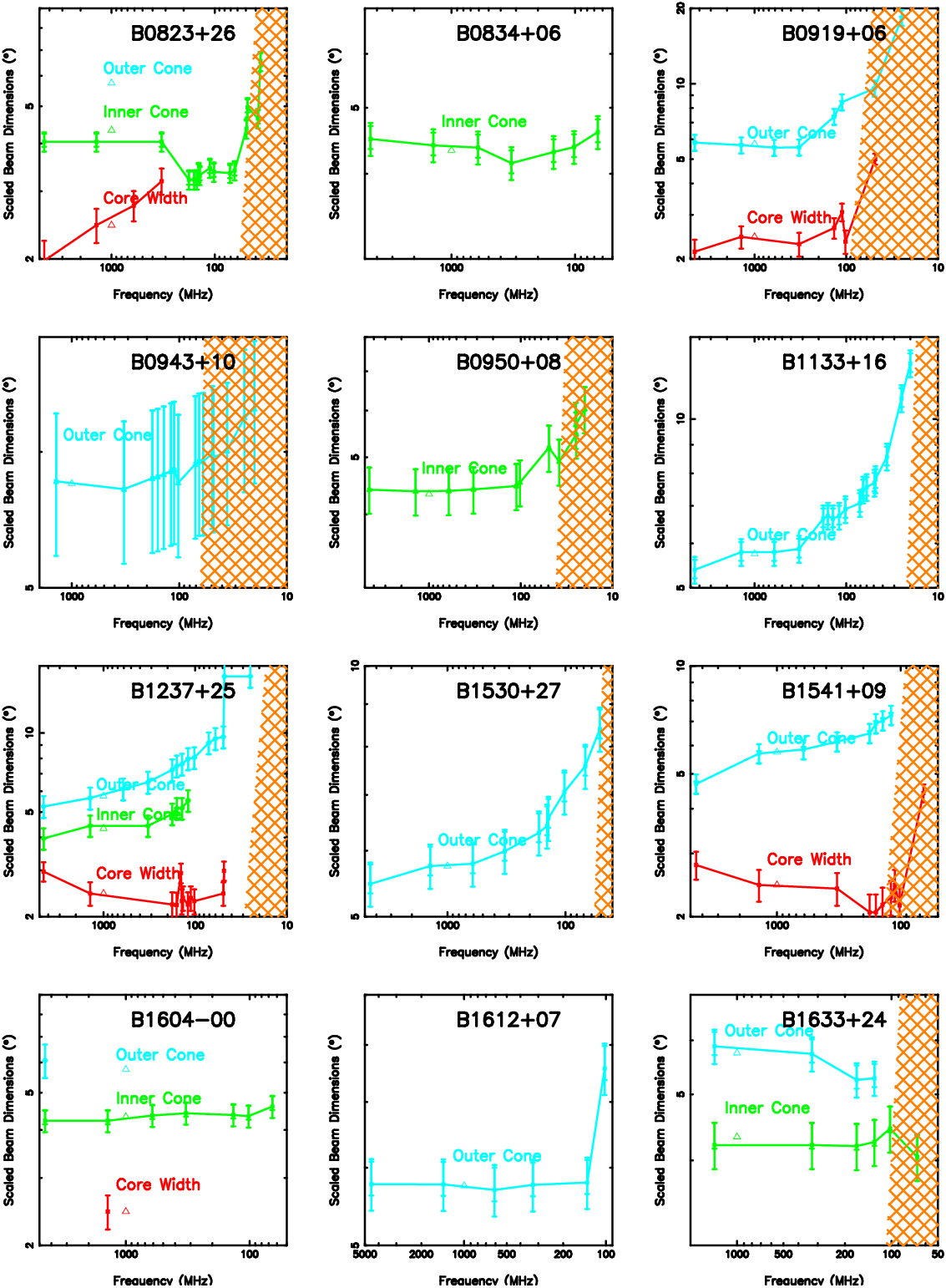}
\caption{Core/double-cone emission-beam model displays for pulsar B0919+06 through J1740+1000.  The \textbf{scaled} angular dimensions of the outer cone (blue), inner cone (green) and core (red) beams are plotted together with errors reflecting 10\% uncertainties in the relevant measurements (see text).  Scattering levels are shown in orange double hashing when measurements are available and yellow single hashing with a 10X boundary when not.}
\label{figA2}
\end{center}
\end{figure*}

\begin{figure*}
\begin{center}
\includegraphics[width=175mm,height=222mm,angle=0.]{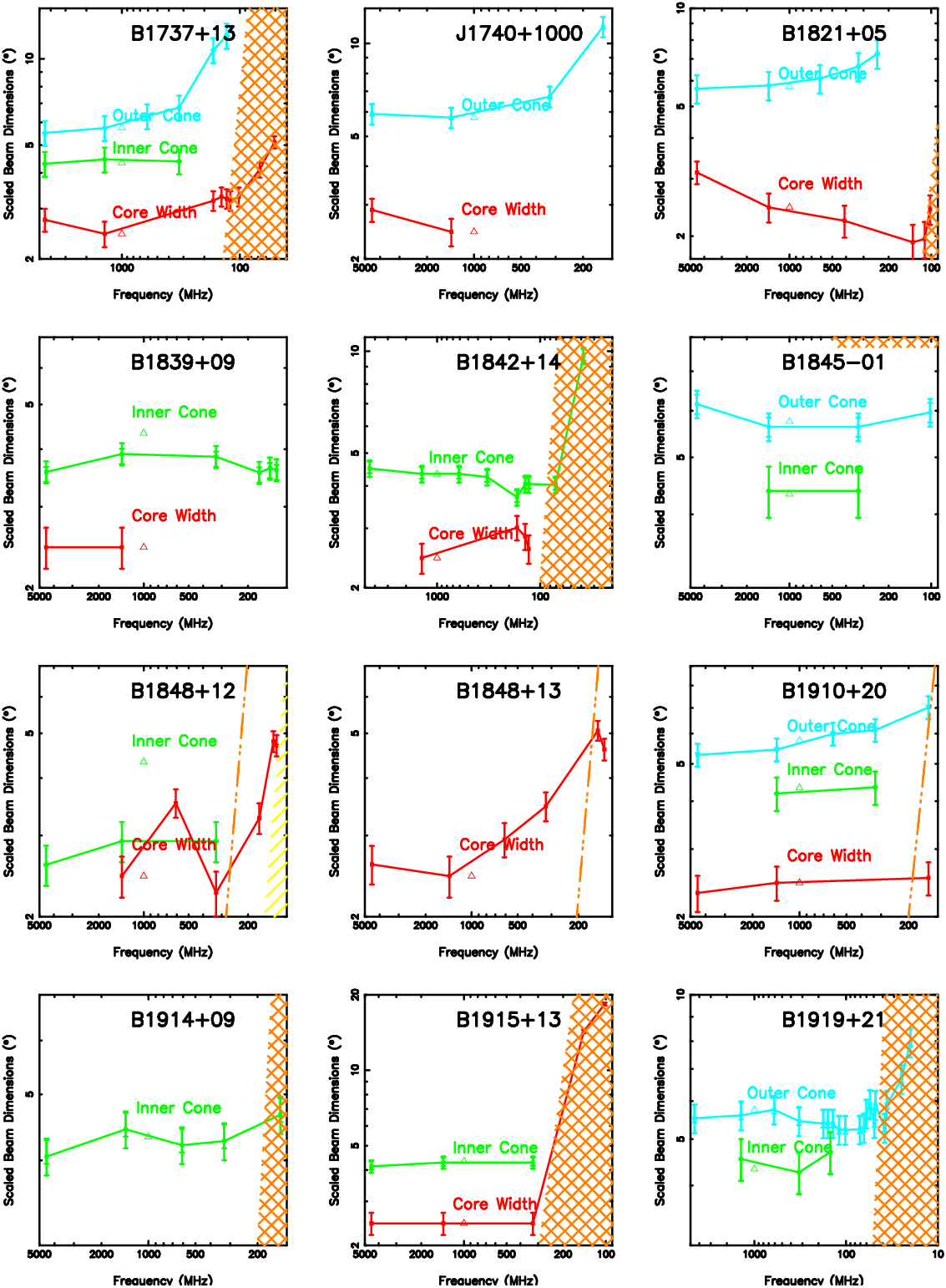}
\caption{Core/double-cone emission-beam model displays for pulsar B1821+05 through B1929+10.  The \textbf{scaled} angular dimensions of the outer cone (blue), inner cone (green) and core (red) beams are plotted together with errors reflecting 10\% uncertainties in the relevant measurements (see text).  Scattering levels are shown in orange double hashing when measurements are available and yellow single hashing with a 10X boundary when not.}
\label{figA3}
\end{center}
\end{figure*}

\begin{figure*}
\begin{center}
\includegraphics[width=175mm,height=222mm,angle=0.]{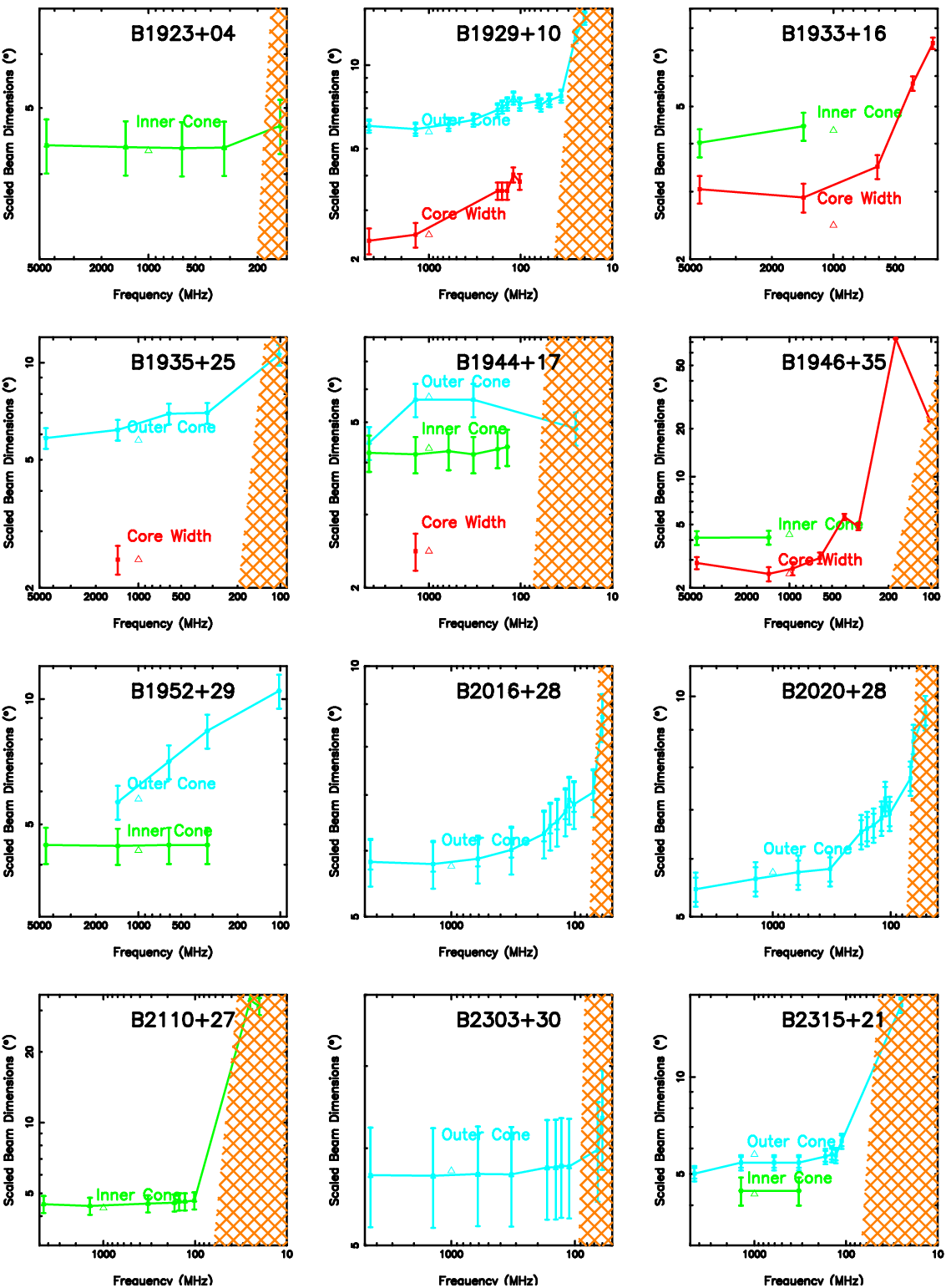}
\caption{Core/double-cone emission-beam model displays for pulsar B1933+16 through B2315+21.  The \textbf{scaled} angular dimensions of the outer cone (blue), inner cone (green) and core (red) beams are plotted together with errors reflecting 10\% uncertainties in the relevant measurements (see text).  Scattering levels are shown in orange double hashing when measurements are available and yellow single hashing with a 10X boundary when not.}
\label{figA4}
\end{center}
\end{figure*}

\begin{table*}
\setlength{\tabcolsep}{3pt}
 \caption{Profile Geometry Information. Given in the first column are pulsar name and profile class. Note that classes with some degree of ambiguity in classification are marked with a question mark, while those with severe ambiguity are marked with two question marks. The next column give the magnetic colatitude $\alpha$, impact angle $\beta$, and sweep rate $R$ as measured at 1-GHz. Subsequent columns give the core width $W_{c}$, inner and outer conal widths, $W_{i}$ and $W_{o}$, and scaled inner and outer beam radii, $\rho_{i}$ and $\rho_{o}$ at each frequency profile information was available at. Due to the difficulty of identifying the inner and outer cones at low frequencies, the conal width $W_{i,o}$ and scaled conal beam radii $\rho_{i,o}$. Lastly, at bottom are listed decametric core widths.}
  \begin{tabular}{lc|ccc|ccccc|ccccc|ccccc|cc}
  \toprule
    Pulsar &  Class & $\alpha$ & $R$ & $\beta$ &  $W_{c}$ & $W_i$ & $\rho_i$ & $W_o$  & $\rho_o$ & $W_{c}$ & $W_i$ & $\rho_i$ & $W_o$  & $\rho_o$ & $W_{c}$ & $W_i$ & $\rho_i$ & $W_o$  & $\rho_o$ & $W_{i,o}$  & $\rho_{i,o}$ \\
  &   & (\degr) & (\degr) & (\degr/\degr) & (\degr) & (\degr) & (\degr) & (\degr) & (\degr) & (\degr) & (\degr) & (\degr) & (\degr) &(\degr) & (\degr) & (\degr) & (\degr) & (\degr) & (\degr) & (\degr) & (\degr) \\
  \midrule
  & & \multicolumn{3}{c|}{(1-GHz Geometry)} & \multicolumn{5}{c|}{(4.5-GHz Dimensions)} & \multicolumn{5}{c|}{(1-GHz Dimensions)} & \multicolumn{5}{c|}{(100-MHz Dimensions)} & \multicolumn{2}{c}{($<$100 MHz)} \\
  \midrule
  \midrule
  B0301+19 & D & 40 & -17 & +2.2 &  --- &  --- &  --- & 10.9 & 4.2 &  --- &  --- &  --- & 13.0 & 4.8 &  --- &  --- &  --- & 19.8 & 7.2 & 28.8 & 9.7 \\
B0523+11 & cQ/M? & {\bf 83} & -9.5 & -6.0 &  --- &  --- &  --- & 14.9 & 9.5 &  --- &  --- &  --- & 14.9 & 9.5 &  --- &  --- &  --- & 17.7 & 10.6 &  --- &  --- \\
B0525+21 & D & 21 & +36 & +0.6 &  --- &  --- &  --- & 13.8 & 2.6 &  --- &  --- &  --- & 15.9 & 2.9 &  --- &  --- &  --- & 22.0 & 4.0 & 26.6 & 4.9 \\
J0538+2817 & T/M? & {\bf 46} & -+5 & +8.3 & +8.0 & 25.0 & 12.6 &  --- &  --- & 9 & 19.7 & 11.2 & 34.0 & 15.4 & 0 &  --- &  --- &  --- &  --- &  --- &  --- \\
B0540+23 & St & {\bf 38} & -3.5 & -10.0 & {$\sim$}9 &  --- &  --- & 17.5 & 11.1 & 8.1 &  --- &  --- & ~20 & 11.4 & 11.8 &  --- &  --- & 22.0 & 11.6 &  --- &  --- \\
\\[-1pt] 
B0609+37 & T & {\bf 50} & +24 & 1.8 & +6.2 & 17.2 & 6.9 &  --- &  --- & 5.9 & 17 & 6.8 &  --- &  --- &  --- & 30.9 & 13.2 &  --- &  --- & 48.0 & 18.5 \\
B0611+22 & St & {\bf 35} & +5 & +6.7 & +4.1 & 8.2 & 7.1 &  --- &  --- & 7.4 & 11.1 & 7.5 &  --- &  --- &  --- & 22.5 & 9.7 &  --- &  --- &  --- &  --- \\
B0626+24 & M/cQ & 63 & -12? & +4.3 &  --- & 8.2 & 5.6 & 15.5 & 8.2 &  --- & 10 & 6.2 & {$\sim$}16 & 8.4 &  --- & 15.2 & 8.1 & 17.1 & 8.8 &  --- &  --- \\
J0627+0649 & T?? & 28 & -3? & +8.6 &  --- &  --- &  --- & 17.0 & 9.7 & 9 &  --- &  --- & {$\sim$}17 & 9.7 & 0 &  --- &  --- &  --- &  --- &  --- &  --- \\
J0631+1036 & cQ & 52 & -6.8 & -6.7 & +0.0 & 8.3 & 7.3 & 21 & 10.3 & 0 & 8.3 & 7.3 & 22.5 & 10.7 &  --- &  --- &  --- & 52.7 & 20.8 &  --- &  --- \\
\\[-1pt] 
B0656+14 & T & {\bf 19} & -4? & +4.7 & +12.7 & 26.4 & 6.7 &  --- &  --- & {$\sim$}12 & 30 & 7.2 &  --- &  --- &  --- & 26.3 & 6.7 &  --- &  --- &  --- &  --- \\
B0751+32 & D & 26 & +25 & +1.0 &  --- &  --- &  --- & 20.6 & 4.7 &  --- &  --- &  --- & 21.1 & 4.8 &  --- &  --- &  --- & 24.8 & 5.6 &  --- &  --- \\
B0823+26 & St & {\bf 84} & +18 & +3.3 & +2.7 & 9 & 5.5 &  --- &  --- & 3.38 & 9 & 5.5 & {$\sim$}14 & 7.7 &  --- & 6.6 & 4.6 &  --- &  --- & 16.8 & 9.0 \\
B0834+06 & D & 50 & +17.0 & +2.6 &  --- & 7.8 & 4.0 &  --- &  --- &  --- & 7.5 & 3.9 &  --- &  --- &  --- & 7.4 & 3.9 &  --- &  --- & 8.1 & 4.1 \\
B0919+06 & T & {\bf 53} & +6 & +7.6 & +4.1 &  --- &  --- & 11 & 8.9 & {$\sim$}4.7 & 10 &  --- & 10 & 8.7 & 4.5 &  --- &  --- & 10.0 & 28.0 & 65.0 & 28.0 \\
\\[-1pt] 
B0943+10 & Sd & 11.5 & -2.1 & -5.4 &  --- &  --- &  --- &  --- &  --- &  --- &  --- &  --- & {$\sim$}11 & 5.5 &  --- &  --- &  --- & 16.7 & 5.6 & 37.0 & 6.1 \\
B0950+08 & Sd? & 12 & -1.4 & +8.5 &  --- & 14.3 & 8.8 &  --- &  --- &  --- & 12.2 & 8.7 &  --- &  --- &  --- & 21.2 & 9.0 &  --- &  --- & 35.2 & 9.8 \\
B1133+16 & D & 46 & +10 & +4.1 &  --- &  --- &  --- & 7.3 & 4.9 &  --- &  --- &  --- & 9.0 & 5.3 &  --- &  --- &  --- & 12.3 & 6.3 & 39.0 & 11.5 \\
B1237+25 & M & {\bf 53} & -150 & -0.3 & +3.2 & 8.4 & 3.4 & 11.1 & 4.5 & {$\sim$}2.6 & 9.4 & 3.8 & 12.0 & 4.8 & 2.4 & 11.7 & 4.7 & 17.0 & 6.9 & 35.0 & 14.0 \\
B1530+27 & Sd & 30 & +5.8 & 4.9 &  --- &  --- &  --- & 5.5 & 5.2 &  --- &  --- &  --- & 8.3 & 5.4 &  --- &  --- &  --- & 13.9 & 6.7 & 23.1 & 7.9 \\
\\[-1pt] 
B1541+09 & T & {\bf 7.4} & -2.2 & -3.4 & {$\sim$}25 &  --- &  --- & 91.7 & 5.4 & ~22 &  --- &  --- & 125 & 6.6 & 19.3 &  --- &  --- & 180 & 8.4 &  --- &  --- \\
B1604-00 & T & {\bf 48} & -8? & +5.3 &  --- & 9.8 & 6.5 & 20 & 9.3 & 5.1 & 9.8 & 6.5 &  --- &  --- &  --- & 10.6 & 6.7 &  --- &  --- & 12.2 & 7.1 \\
B1612+07 & Sd & 24 & -4.6 & +5.1 &  --- &  --- &  --- & 6.2 & 5.3 &  --- &  --- &  --- & 6.1 & 5.3 &  --- &  --- &  --- & 21.0 & 6.9 &  --- &  --- \\
B1633+24 & cT/Q & 19 & -4 & -5.3 &  --- &  --- &  --- &  --- &  --- &  --- & 20 & 6.0 & 47 & 8.4 &  --- & 20.9 & 6.1 & 38.5 & 7.5 & 15.9 & 5.8 \\
B1737+13 & M & {\bf 36} & -12 & +2.8 & +5.2 & 12.6 & 4.8 & 17.8 & 6.2 & {$\sim$}4.6 & 13.3 & 5.0 & 18.7 & 6.4 & 6.2 &  --- &  --- &  --- &  --- &  --- &  --- \\
\\[-1pt] 
J1740+1000 & St & {\bf 23} & +1.8 & 12.7 & +18.5 &  --- &  --- & 33.1 & 15.1 & 15.7 &  --- & 13.3 & 29.9 & 14.7 &  --- &  --- &  --- & 110 & 28.9 &  --- &  --- \\
B1821+05 & T & {\bf 31} & -10 & -2.9 & +7.0 &  --- &  --- & 23.8 & 6.5 & {$\sim$}5.5 &  --- &  --- & 24.5 & 6.7 & 5.4 &  --- &  --- &  --- &  --- &  --- &  --- \\
B1839+09 & T & {\bf 83} & -19 & +3.0 & +4.0 & 9.9 & 5.8 &  --- &  --- & {$\sim$}4.0? & 11.2 & 6.3 &  --- &  --- &  --- & 10.1 & 5.8 &  --- &  --- &  --- &  --- \\
B1842+14 & T/St & {\bf 71} & +12 & 4.5 &  --- & 12.1 & 7.4 &  --- &  --- & {$\sim$}4.2 & 11.4 & 7.1 &  --- &  --- & 4.5 & 10.1 & 6.6 &  --- &  --- & 31.7 & 15.8 \\
B1845-01 & cT & {\bf 39} & +8 & +4.5 &  --- &  --- &  --- & 18.5 & 7.6 &  --- & 9 & 5.4 & 16 & 6.9 &  --- &  --- &  --- & 17.6 & 7.3 & 0.0 & 0.0 \\
\\[-1pt] 
B1848+12 & St/T & {\bf 63} & 36 & 1.4 &  --- & 4.2 & 2.4 &  --- &  --- & {$\sim$}2.5 & 5 & 2.7 &  --- &  --- & 4.8 &  --- &  --- &  --- &  --- &  --- &  --- \\
B1848+13 & St & {\bf 44} &  --- &  --- & +6.4 &  --- &  --- &  --- &  --- & {$\sim$}6 &  --- &  --- &  --- &  --- & 11.3 &  --- &  --- &  --- &  --- &  --- &  --- \\
B1910+20 & M & {\bf 29} & +18 & 1.5 & +3.2 &  --- &  --- & 12.9 & 3.5 & {$\sim$}3.4? & 9.5 & 2.8 & 13.4 & 3.6 & 3.5 &  --- &  --- & 18.0 & 4.7 &  --- &  --- \\
B1914+09 & D? & 52 & -7 & -6.5 &  --- & 11.6 & 7.8 &  --- &  --- &  --- & 14.9 & 8.5 &  --- &  --- &  --- &  --- &  --- & 8.1 & 7.1 &  --- &  --- \\
B1915+13 & St & {\bf 68} & -8 & +6.6 & 6? & 14 & 9.4 &  --- &  --- & 6.0 & 15 & 9.7 &  --- &  --- & 45 &  --- &  --- & 8.1 & 7.7 &  --- &  --- \\
\\[-1pt] 
B1919+21 & cQ & 45 & -11 & -3.7 &  --- &  --- &  --- & 8.9 & 4.8 &  --- & 4 & 3.9 & 9.2 & 4.8 &  --- &  --- &  --- & 7.7 & 4.5 & 10.3 & 5.1 \\
B1923+04 & Sd/cT? & 25 & +6 & +4.0 &  --- & 5.9 & 4.3 &  --- &  --- &  --- & 5.5 & 4.2 &  --- &  --- &  --- &  --- &  --- & 8.1 & 4.4 &  --- &  --- \\
B1929+10 & T/M? & {\bf 88} & -6?? & +10.5 & +4.9 &  --- &  --- & 14.4 & 12.7 & 5.15 &  --- &  --- & 13.2 & 12.4 & 8.0 &  --- &  --- & 24.4 & 16.0 & 25.0 & 16.3 \\
B1933+16 & St & {\bf 55} & -39 & -1.2 & +6.2 & 16.3 & 6.7 & 0.0 & 0.0 & {$\sim$}5.9 & 18 & 7.4 &  --- &  --- &  --- &  --- &  --- &  --- &  --- &  --- &  --- \\
B1935+25 & T? & {\bf 77} & -9 & +6.2 &  --- &  --- &  --- & 23.2 & 13.0 & {$\approx$}6? &  --- &  --- & 25 & 13.8 &  --- &  --- &  --- & 46.2 & 23.6 &  --- &  --- \\
\\[-1pt] 
B1944+17 & cT/cQ & {\bf 4.7} & +0.8 & 6.3 &  --- & 16.2 & 6.4 & 38.7 & 6.7 & {$\approx$}45? & 10 & 6.3 & {$\sim$}95 & 8.5 &  --- &  --- &  --- & 31.1 & 6.6 &  --- &  --- \\
B1946+35 & St/T & {\bf 35} & +16 & 2.1 & +5.9 & 14.9 & 4.9 &  --- &  --- & 5.0 & 15 & 4.9 &  --- &  --- & 46 &  --- &  --- & 31.1 & 9.4 &  --- &  --- \\
B1952+29 & M/cQ? & {\bf 43} & +38 & -1.0 &  --- & 20 & 6.8 &  --- &  --- &  --- & 19.9 & 6.8 & 25.5 & 8.7 &  --- &  --- &  --- & 47.5 & 16.0 &  --- &  --- \\
B2016+28 & Sd & 39 & +5 & +7.2 &  --- &  --- &  --- & 8.6 & 7.8 &  --- &  --- &  --- & 8.2 & 7.7 &  --- &  --- &  --- & 16.6 & 9.1 & 27.0 & 11.6 \\
B2020+28 & D/T? & 88 & +8 & +7.2 &  --- &  --- &  --- & 11.9 & 9.3 &  --- &  --- &  --- & 12.8 & 9.6 &  --- &  --- &  --- & 18.8 & 11.8 & 29.1 & 16.2 \\
\\[-1pt]
B2110+27 & Sd & 48 & -11 & +3.9 &  --- & 3.5 & 4.1 &  --- &  --- &  --- & 2.9 & 4.0 &  --- &  --- &  --- & 4.5 & 4.2 &  --- &  --- &  --- &  --- \\
B2303+30 & Sd & 20.5 & +4.5 & +4.5 &  --- &  --- &  --- & 4.4 & 4.5 &  --- &  --- &  --- & 4.3 & 4.5 &  --- &  --- &  --- & 6.3 & 4.6 & 12.4 & 5.1 \\
B2315+21 & cT/cQ? & 88 & +17 & +3.4 &  --- &  --- &  --- & 4.9 & 4.2 &  --- & 3 & 3.7 & {$\sim$}6 & 4.5 &  --- &  --- &  --- & 8.1 & 5.3 &  --- &  --- \\

  \bottomrule
  \end{tabular}
\label{tabA3}
Decametric core widths: B0919+06, 9.6\degr\ at 48.4 MHz; B1237+25, 7.8\degr, 8.4\degr\ at 71, 48 MHz; B0919+06, 3.2\degr\ at 48.4 MHz; B1541+09, 39.6\degr\ at 65.4 MHz; 
\end{table*}

\begin{figure}
\begin{center}
\includegraphics[width=65mm,angle=-90.]{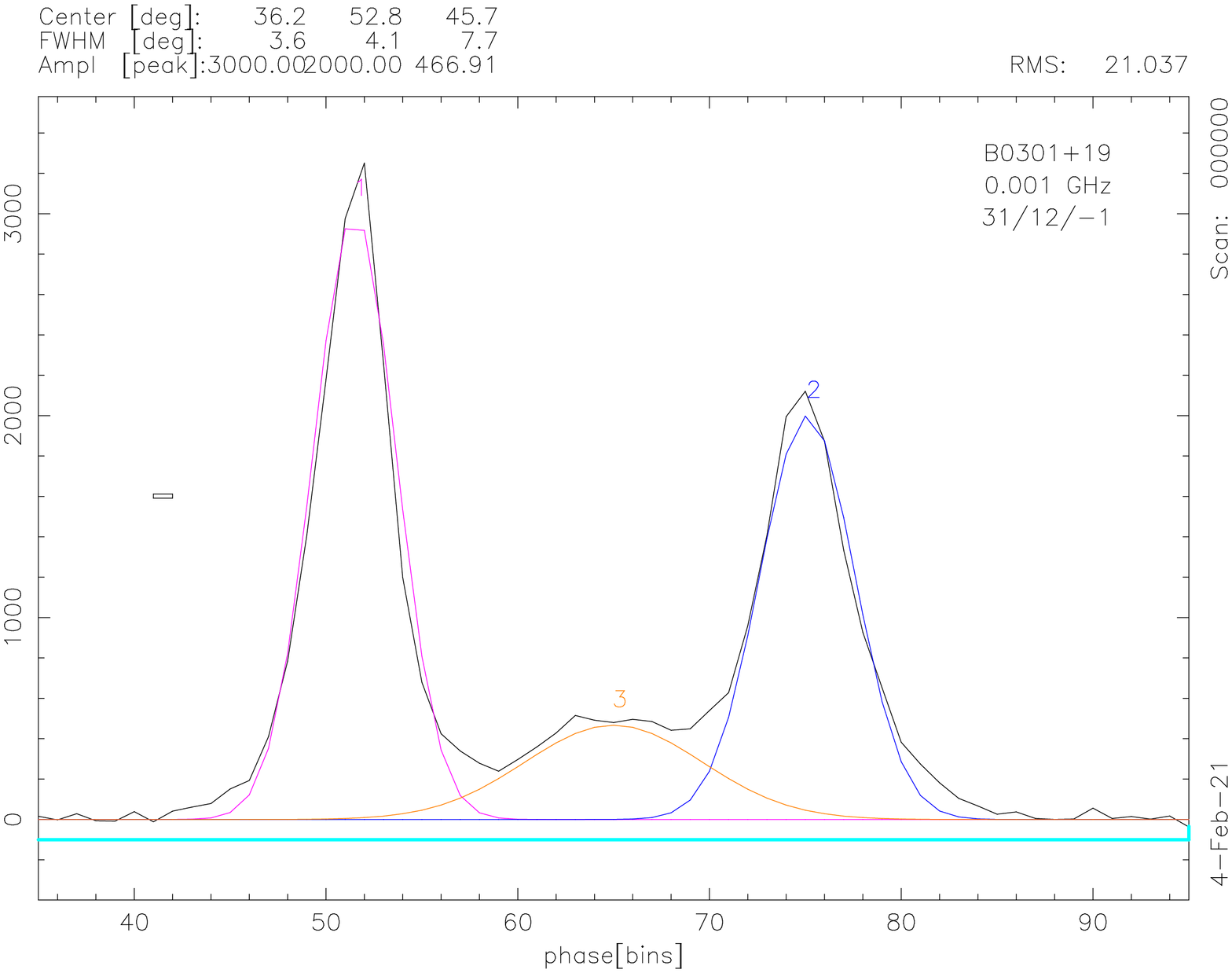}
\caption{B0301+19: The BKK16+ 129-MHz observation is typical of all the PRAO and LOFAR profiles in showing a central component.  The plot shows a fit to this profile showing its three components.  The core width is probably overestimated due to not taking into account inner conal power.}
\label{figA101}
\end{center}
\end{figure}
\noindent\textit{\textbf{B0301+19}}: This well-known conal double pulsar had usually been regarded as having an inner cone \citep{mitra2002}, but the low-frequency observations show that this is incorrect.  A weak core component with an expected width of some 3.2\degr\ is clearly seen in single pulses and at low frequencies \citep{young}---\citep[see also][]{JKMG08}---but its width is difficult to determine accurately.  Core widths of 6.5\degr\ and 7.2\degr\ at 168 and 129 MHz respectively, are upper limits as the features are conflated with the conal components---and in the case of the fitting effort (see Fig.~\ref{figA101}), no inner conal power can be modeled that would constrain the core width.  Overall, the core width could be relatively constant down to 100 MHz or so.  
\vskip 0.1in

\noindent\textit{\textbf{B0523+11}}: The pulsar has been regarded as having an outer cone because some observations hint at an additional set of inner conal components and perhaps a core \citep[see also][]{JKMG08}. However, scattering makes measurements below 100 MHz impossible, so we cannot know whether the usual outer conal RFM is seen.
\vskip 0.1in

\begin{figure}
\begin{center}
\includegraphics[width=65mm,angle=-90.]{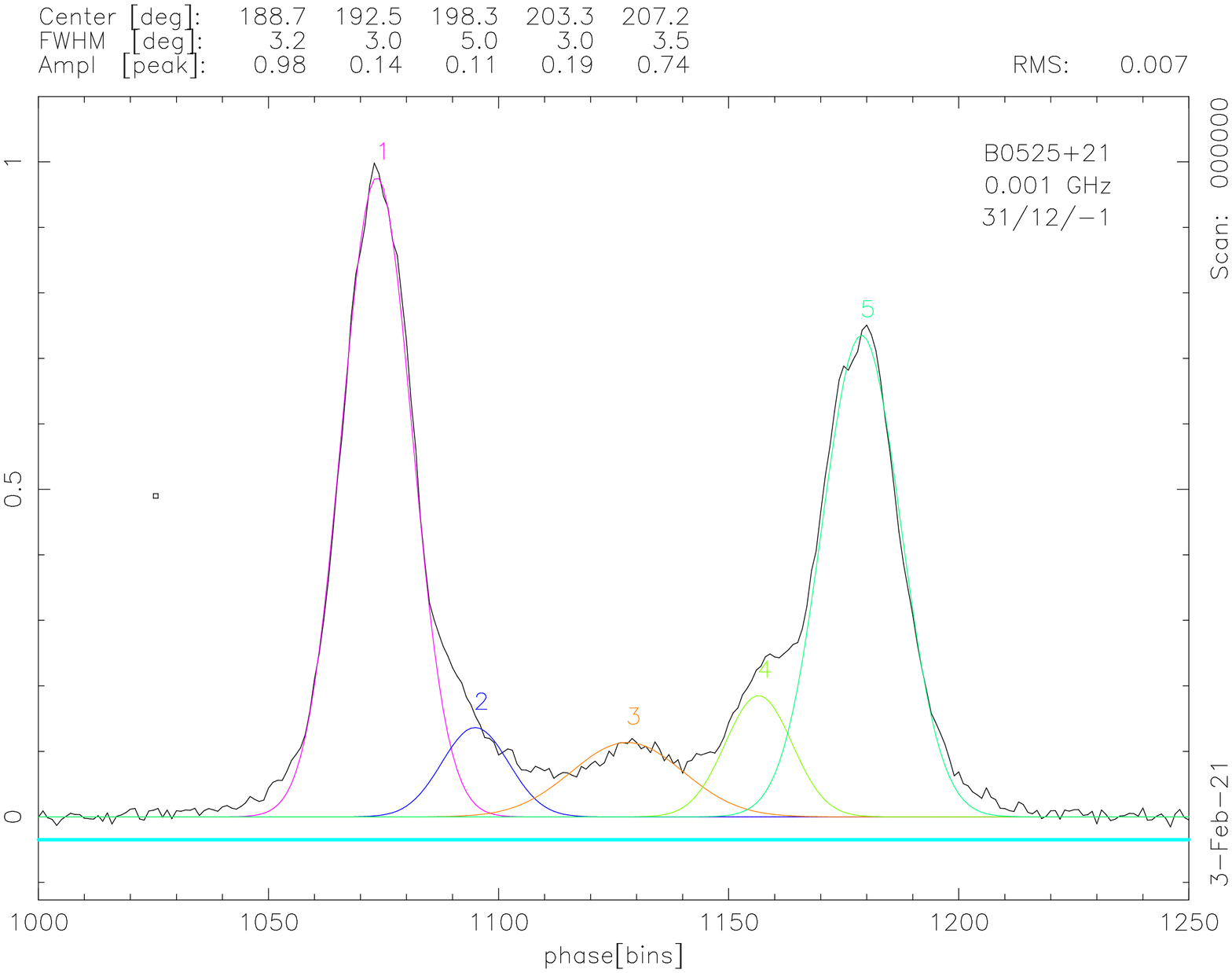}
\caption{B0525+21: The PHS16 149-MHz observation shows inflections corresponding to the inner conal components as well as a central core feature.  The image shows the results of fitting 5 Gaussian components to this profile.}
\label{figA102}
\end{center}
\end{figure}
\noindent\textit{\textbf{B0525+21}}: A classic outer conal double profile  is present in this profile \citep{mitra2002}.  However, single pulses corresponding to the core and inner cone are seen at meter wavelengths \citep{young}.  All the availible LOFAR profiles show a hint of the central core, in particular the 149-MHz (PHS16) profile.  Fig~\ref{figA102} shows the results of Gaussian fitting to the 149-MHz PHS16 profile. 
\vskip 0.1in

\noindent\textit{\textbf{J0538+2817}}: The three OMR19 profiles are all we have to go on, and it is not clear that their structures provide a compatible interpretation, especially that at 4.3 GHz.  The 1.4-GHz and 327-MHz profiles may have five components, but the central component is not clearly marked as a core, and the PPA traverse steepens on the trailing edge as if from aberration/retardation.  An aspirational model is possible only if the PPA rate steepens to some --5/degr/\degr, and this is what we report.  
\vskip 0.1in

\noindent\textit{\textbf{B0540+23}}: The asymmetric profile of PSR B0540+23 has made it perennially difficult to classify.  However, its main feature seems to be core, with a hint of a cone at high frequency that becomes more prominent at 327 MHz on the trailing side of the profile \citep[see also][]{JKMG08, Olszanski2021}.  The LOFAR observations show progressive scattering broadening with a width of perhaps 10\degr\ at 178 MHz, so any conal contribution is obscured and the core width value dominated by scattering.   
\vskip 0.1in

\noindent\textit{\textbf{B0609+37}}: This pulsar seems to be an inner-cone triple that may well exhibit modal changes in the strength of its conal components.  The core is only discernible at high frequency.
\vskip 0.1in

\noindent\textit{\textbf{B0611+22}}: This pulsar has what seems to be a \textbf{S$_t$} profile at high frequency. Its emission comes in complex bursts as has been studied by \citet{seymour}.  Substantial scattering is expected at 200 MHz and below. See also \citet{JKMG08,Olszanski2021}.  

\begin{figure}
\begin{center}
\includegraphics[width=75mm,angle=-90.]{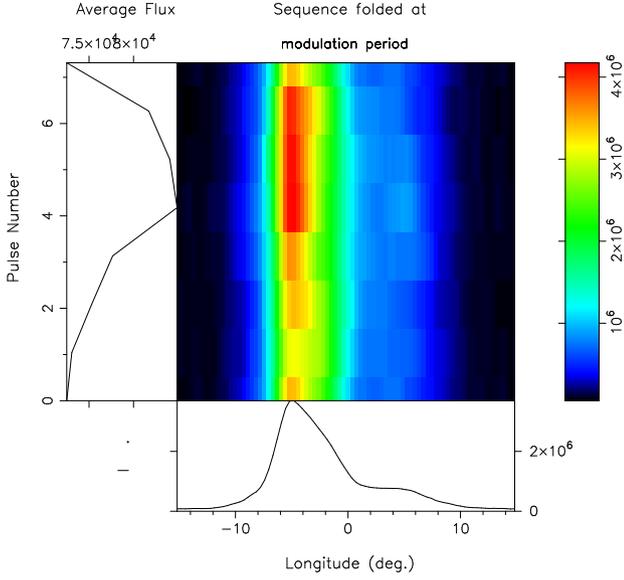}
\caption{B0626+24: A 7.3-$P$ cycle  is seen in many pulse intervals.  Here a 512-pulse section is folded on this cycle to show how different parts of the profile are modulated. Color is used to represent the intensity of emission. The first and last pulse displayed are the same, as the beginning and end of the cycle overlap. }
\label{figA103}
\end{center}
\end{figure}
\vskip 0.1in

\noindent\textit{\textbf{B0626+24}}:  Two cones are seen at high frequency in this pulsar, and a subpulse modulation cycle of 7.3-$P$ as shown in Fig.~\ref{figA103}. The pulsar is visible down to 100 MHz, with clearly discernible scattering.
\vskip 0.1in

\noindent\textit{\textbf{J0627+0649}}:  Again, OMR19 is the only source for this pulsar.  The 1.4- and 4.5-GHz profiles have similar forms, while the 327-MHz profile is so different that it is guesswork to compare it.  The 1.4-GHz suggests a five component profile, but the PPA traverse is so shallow that an inner conal model fails.  We therefore model it as a core-cone triple beam system, and the extension to the low frequency remains questionable.  
\vskip 0.1in

\noindent\textit{\textbf{J0631+1036}}: This pulsar presents a rare example of a four-component pulsar.  \citet{teixeira} studied the pulsar and worked out the geometry that we use here to model its profiles.  The authors outline issues with accommodating the inner cone, but the dimension do satisfy the double cone model.
\vskip 0.1in

\noindent\textit{\textbf{B0656+14}}:  In spite of this pulsar's unusual properties (\eg \citet{weltevrede2006a,weltevrede2006b}) it seems to have a core/inner-cone profile, such that the conal emission modulates the edges of the core beam.  The two are apparently conflated for the most part, but a hint of this configuration is seen in the \citet{gould} profiles as well as the tripartite form of the LOFAR 149-MHz profile.  See also \citet{Olszanski2021}.
\vskip 0.1in

\noindent\textit{\textbf{B0751+32}}: A conal double \textbf{D} profile is seen in PSR B0751+32.  Fluctuation spectra show a hint of a 6-$P$ cycle \citep{weltevrede2007}, however frequent nulls of about 10 periods make this difficult to measure.  The pulsar has been regarded as having an inner cone, but the significant RFM could make it an outer cone were its RFM discernible below 100 MHz.  The BKK19 60-MHz profile is apparently highly scattered.
\vskip 0.1in

\noindent\textit{\textbf{B0823+26}}:  The bright mode main-pulse (MP) profile width at higher frequencies in this pulsar is dominated by the core component, but a highly conflated conal features can also be detected \citep{rankin_olszanski2020, rankin1997} using single-pulse analyses.  See also \citet{basu2018}.  Down to and below 100 MHz the MP width increases---perhaps because the core width increases---but more likely because the core emission becomes less dominant and the inner cone (or even the outer) contributes more importantly to the width.  That it falls below the expected inner cone width argues that core contributions still remain strong. In the decametric band, conal emission may dominate the MP profile width.  Zakharenko \etal\ (2013) are able to measure MP profile widths in the 25 and 20-MHz bands and estimate the scattering at only 25\degr\ at 25 MHz.  Thus, meaningful profile widths can be measured, though it remains unclear what type of emission produces them.  The beam model plot for the pulsar then requires some comment:  the core width narrows at 4.5 GHz, perhaps because a modal half is less prominent.  Its 1-GHz width seems to accurately reflect the polar cap width, and the 327-MHz width includes some conal contributions.  The inner and even outer conal contributions to the width can be roughly discerned at the above three frequencies \citep{rankin_olszanski2020}, but no such analysis is available for the lower frequency profiles.  The narrow putative inner cone widths at around 100 MHz then probably reflect the admixture of core and conal power.
\vskip 0.1in

\begin{figure}
\begin{center}
\includegraphics[width=75mm,angle=0.]{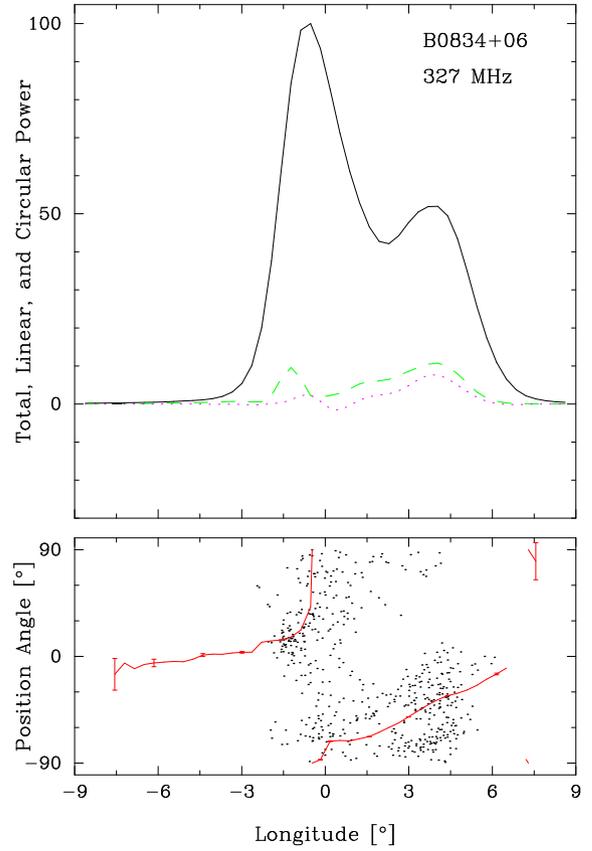}
\caption{B0834+06: In the top panel is plotted the total intensity (black solid line), in addition to the linearly polarized (dashed green line), and circularly polarized (purple dotted line) intensities. The bottom panel shows a histogram of polarization position angles with the average PPA track overlaid (red line). Note that B0834+06's profile has a discernible level of polarization and well defined PPA out to half of the profile width prior to the leading component. B0834+06 is well-classified as an inner cone and inner cones typically possess edge dopolarization. The lack of it in this pulsar is unsual.}
\label{figA104}
\end{center}
\end{figure}
\noindent\textit{\textbf{B0834+06}}: One of the four original Cambridge pulsars, PSR B0834+06 has a classic narrow inner-cone double profile that shows very little width increase with wavelength down to some 60 MHz \citep{mitra2002}. It shows a prominent, nearly even-odd subpulse modulation \citep{rankinwright2007} supporting its conal character.  Linear power appearing in organized subpulses occurs much earlier than the profile onset as shown in Fig.~\ref{figA104}; several prominent inner conal double profiles show this effect.  Zakharenko \etal (2013) measure a scattering width of 17\degr\ at 25 MHz, so little of the intrinsic profile structure survives.  
\vskip 0.1in

\noindent\textit{\textbf{B0919+06}}: PSR B0919+06 is famous for its "swooshes" (\citep{bistable,wahl} which conflate the profile as observed; but when these are accounted for it seems to have a usual core/cone \textbf{T} profile \citep[see also][]{JKMG08, Olszanski2021}.  The pulsar's unusual hat-shaped profiles at low frequency (see \citet{hankins2010})) show that the core survives to low frequencies; however, these widths are affected by both the "swooshes" and possible scattering.  The shape is shown clearly in the PHS16's 135-MHz LOFAR profile.  At decameter wavelengths the profiles broaden further, and \citep{Zakharenko2013} find that the 50\degr\ scattering width at 25 MHz is substantially less than that of the profile.  Whether scattering or ``swooshes'' are responsible for this very low frequency width escalation is not clear. 
\vskip 0.1in

\noindent\textit{\textbf{B0943+10}}: This pulsar has a very well-studied outer conal profile that can be observed down to 30 MHz or below \citep[\eg ]{Zakharenko2013}.  It exhibits two modes with different profile forms that complicate modeling; however, for this pulsar these are well understood (\eg \citet{deshpande}).  Note that the 10\% error in the PPA sweep rate gives large fully correlated outer cone beam-radius errors for this pulsar. Scattering is minimal with $t_{\rm scatt}$ at 25 MHz only so 12 ms \citep{Zakharenko2013}. 
\vskip 0.1in

\noindent\textit{\textbf{B0950+08}} An inner cone seems to be present in this pulsar, but the substantial broadening and bifurcation below 100 MHz, suggests that it could have an outer cone. Its interpulse still remains a mystery. Scattering is minimal with $t_{\rm scatt}$ at 25 MHz only 1 ms \citep{Zakharenko2013}. 
\vskip 0.1in

\noindent\textit{\textbf{B1133+16}}: This pulsar has a classic wide double \textbf{D} profile (\eg \citet{mitra2002}).  As one of the brightest pulsars in the northern sky, it has been studied intensively since its discovery, and the modal depolarization of its profile edges can be followed down to the 40 db level \citep{et8}.  In addition, \citet{young} showed that sporadic core emission contributes to this pulsar's intrapulse ``baseline'' region.  Scattering is minimal with $t_{\rm scatt}$ at 25 MHz only 6.5 ms (Zakharenko \etal\ 2013).
\vskip 0.1in

\begin{figure}
\begin{center}
\includegraphics[width=65mm,angle=-90.]{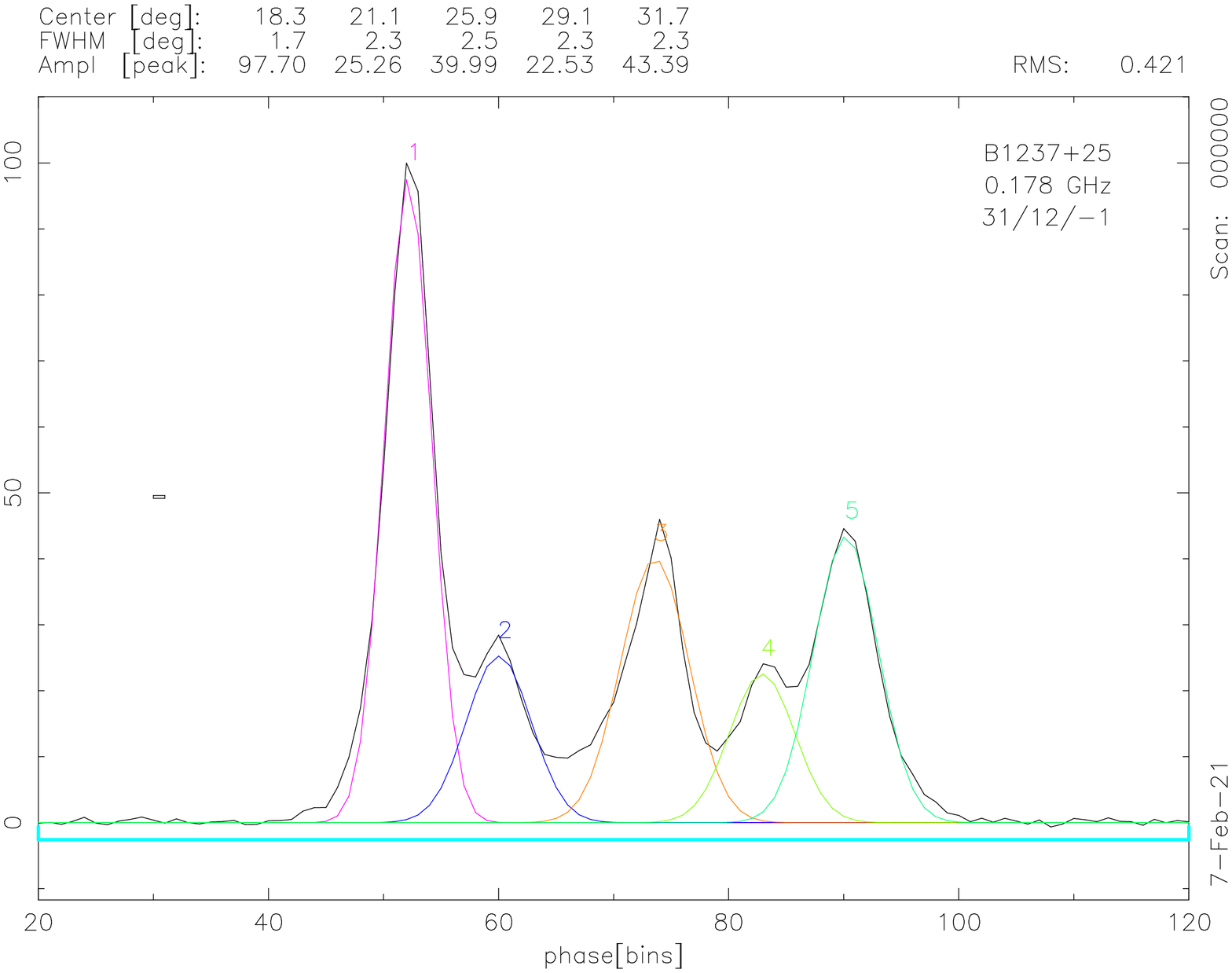}
\includegraphics[width=65mm,angle=-90.]{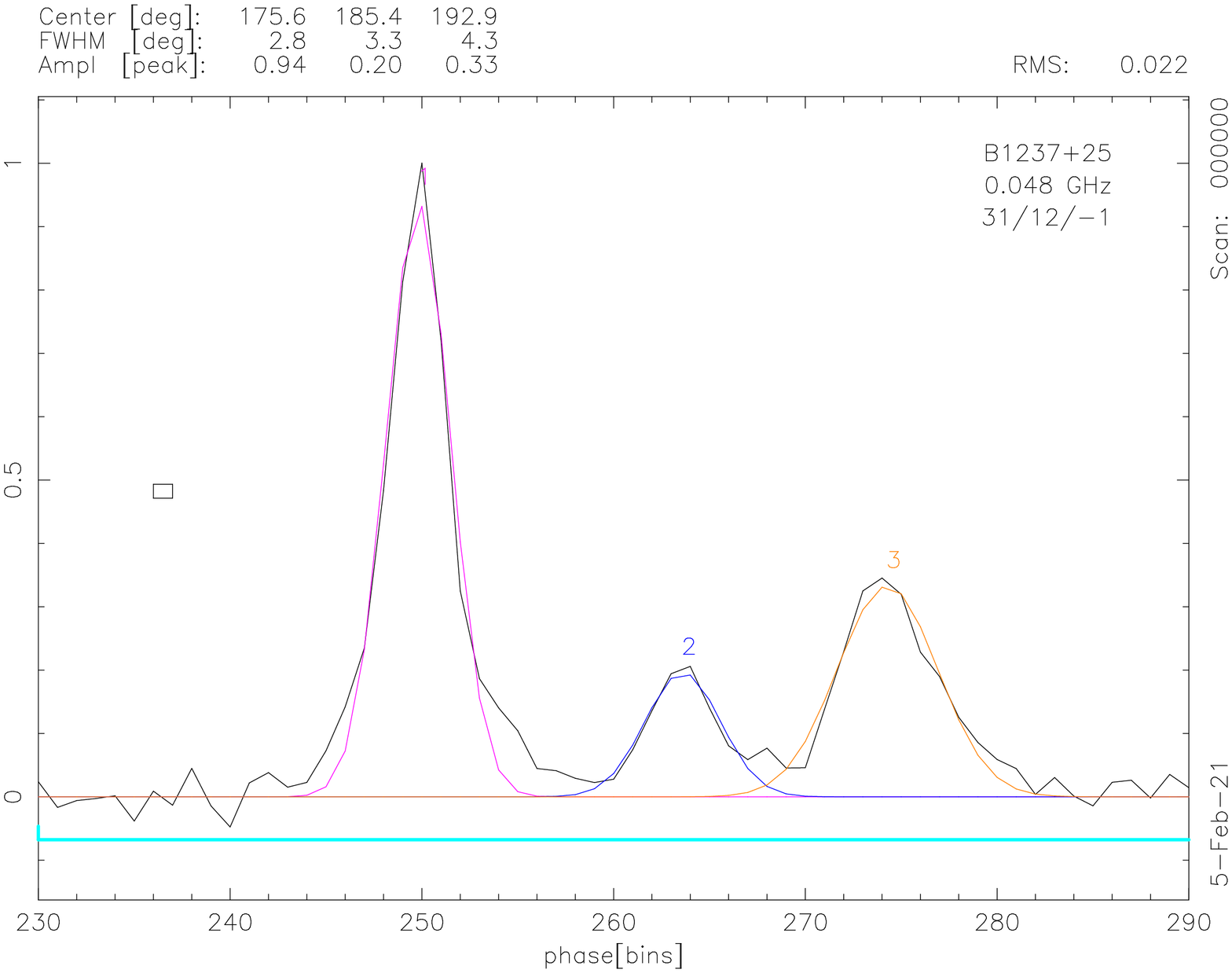}
\caption{B1237+25: Two of the 8 LOFAR profiles we have fitted with Gaussian components to better estimate their core and conal widths: (top) the BKK16+ 178-MHz and (bottom) the 48.4-MHz PHS16 profiles.  The image shows the results of fitting 5 Gaussians to the top profile and 3 to the bottom one.}
\label{figA105}
\end{center}
\end{figure}

\noindent\textit{\textbf{B1237+25}}: This pulsar exhibits its well-known \textbf{M} profile over a broad frequency range (\eg \citet{mitra2002}). Its complex PPA traverse can be understood in detail as indicating a highly central sightline traverse, and its modes exemplify important aspects of both core and conal emission \citep{smith}.  Its small DM and minimal scattering permit meaningful measurements well into the decameter band.  Even its core can be discerned at low frequencies, but measurement accuracy suffers from the core's two parts, a strong trailing and weak leading half as well as some effects from an uncertain proportion of abnormal mode pulses.  We have fitted all eight of the published LOFAR profiles, two of which are given in Fig.~\ref{figA105}.  A number of core width values falling around 2.5\degr\ can be measured from the profiles in BKK16+ and PHS16, where some effort can be made to compensate for its weak early part, but the highest quality are HR10's 2.6\degr\ widths from AO at 111.5 and 49.2 MHz.  Note also the inner cone growth at between 100 and 200 MHz.  Scattering is minimal with $t_{\rm scatt}$ at 25 MHz only 12 ms (Zakharenko \etal\ 2013).
\vskip 0.1in

\begin{figure}
\begin{center}
\includegraphics[width=75mm,angle=-90.]{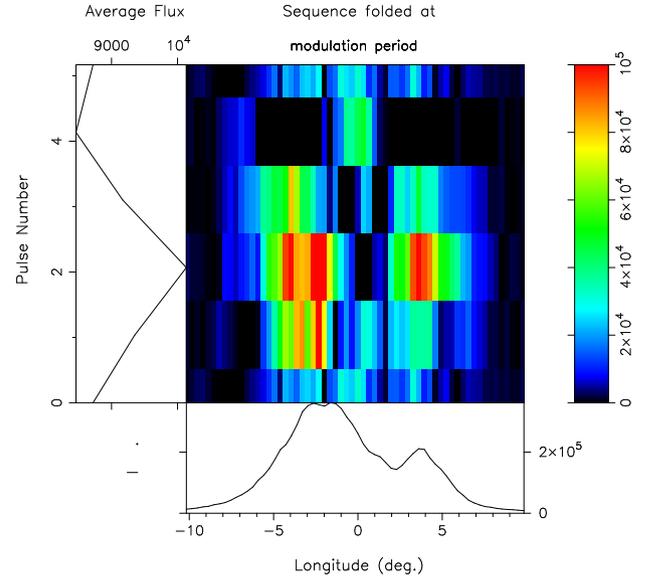}
\caption{B1530+27: The 1024-pulse sequence folded at the 5.2-$P$ cycle shows the typical conal crenellation in both components.}
\label{figA106}
\end{center}
\end{figure}
\vskip 0.1in
\noindent\textit{\textbf{B1530+27}}: PSR B1530+27 shows a very typical outer conal \textbf{S$_d$} profile evolution as well as the very usual conal amplitude modulation as shown in Fig~\ref{figA106}.  The S/N of the \citep{Zakharenko2013} profile is inadequate to trace the profile evolution into the decameter band, though the scattering does not seem to be an obstacle.
\vskip 0.1in

\noindent\textit{\textbf{B1541+09}}: Higher frequency core widths have been overestimated in this pulsar due to conflation with the conal emission. A value of about 20\degr\ suggests an outer conal \textbf{T} structure, and evidence for RFM is seen in 100-MHz profiles  This indicates an $\alpha$ value around 7.5\degr\ as well as a consistent value of the core width down to below 100 MHz where scattering begins to be observed.  We find a 7-5-$P$ cycle in our 327-MHz observation that modulates both the leading and middle component, raising the possibility of a conal triple configuration. However, the quantitative geometry for this structure would require a shallower PPA traverse than is observed.  
\vskip 0.1in

\begin{figure}
\begin{center}
\includegraphics[width=65mm,angle=-90.]{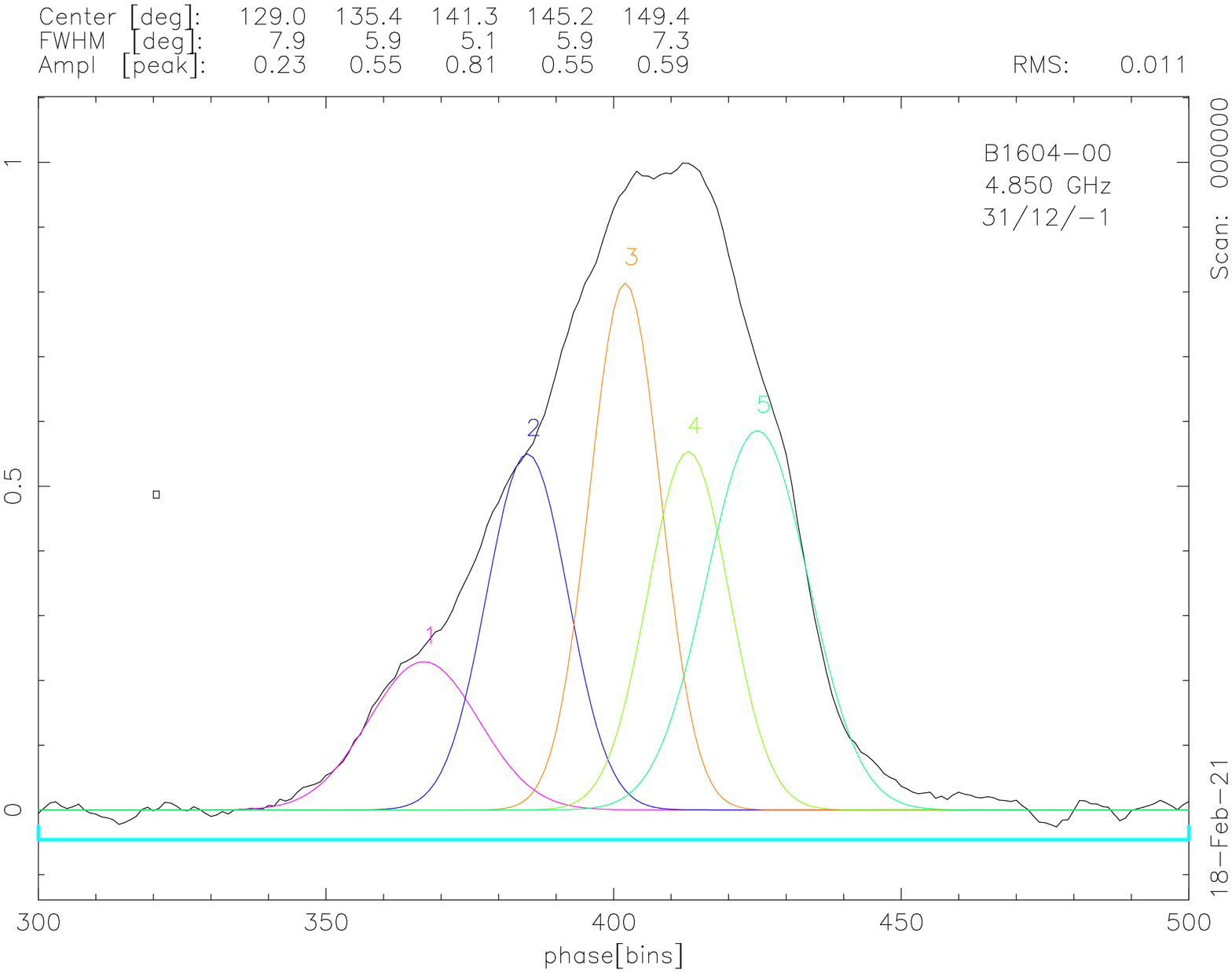}
\caption{B1604--00: The image shows the results of fitting 5 Gaussians to the 4.85-GHz \citep{Kijak1998}}
\label{figA107}
\end{center}
\end{figure}
\noindent\textit{\textbf{B1604--00}}: This pulsar is one of the best examples of inner cones showing little or no RFM or scattering down to about 60 MHz (see \citep{mitra2002}).  However, the profile is broad  at 4.85 GHz, and both that and the 1.4-GHz profile show a structure suggesting an additional outer set of ``outriding'' components.  Gaussian fits to these profiles are shown in Fig.~\ref{figA107}.  The decametric profiles are ``scattered out'' \citep{Zakharenko2013}. 
\vskip 0.1in

\noindent\textit{\textbf{B1612+07}}: The significant RFM in PSR B1612+07 suggests the presence of an outer cone.  Fluctuation spectra show a 5.7-$P$ phase modulation cycle that reflects conal emission. The poor S/N of the \citep{Zakharenko2013} profiles does not permit tracing the profile evolution into the decametric band.
\vskip 0.1in

\begin{figure}
\begin{center}
\includegraphics[width=75mm,angle=-90.]{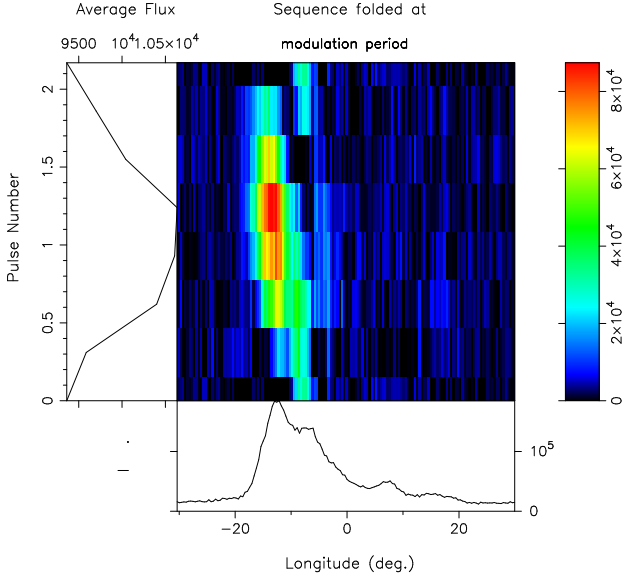}
\caption{B1633+24 shows a very strong nearly even-odd (2.17$P$) in its two strong early components, and some trace of it also modulates the weak trailing components.  The plots gives a folded 1024-pulse sequence with the unvarying ``base'' removed.}
\label{figA108}
\end{center}
\end{figure}
\noindent\textit{\textbf{B1633+24}}:  This \textbf{cT} pulsar (see \citep{hankins1987}) can be observed down to 65 MHz, but unusually, the poor PHS16 profile almost certainly shows only the strong leading components.  That the structure is conal is clear from the strong narrow fluctuation feature at 2.17-$P$, and folded cycle in Fig.~\ref{figA108} shows how the profile is modulated.  The poor S/N of the\citet{Zakharenko2013} profiles does not permit tracing the profile evolution into the decametric band.
\vskip 0.1in

\noindent\textit{\textbf{B1737+13}}: Analyses in vHKK97 show the core clearly at 4-5 degrees, and its width may be conflated at meter wavelengths \citep{force}.  See OMR19 for higher frequency profiles.  
\vskip 0.1in

\noindent\textit{\textbf{J1740+1000}}: This 150-ms energetic pulsar seems to have a core-single \textbf{S$_t$} configuration, first discussed in \citet{Olszanski2019}.  \textbf{Fluctuation spectra computed from our observations showing only red noise (not shown) support this interpretation.}
\vskip 0.1in

\noindent\textit{\textbf{B1821+05}}:  The pulsar seems to have a core-cone triple \textbf{T} configuration.  The fluctuation spectra show nothing other than noise power. Only the core component is seen around 100 MHz.  Its width is narrower here than at higher frequencies as if a leading part of it is missing as for B1237+25---see the \citet{hankins2010} 430-MHz profile.  An outer cone is indicated by the core width even though no RFM is observed.
\vskip 0.1in

\noindent\textit{\textbf{B1839+09}}:  A central core component with conal ``outriders'' is seen at high frequency in this pulsar; whereas  only two conal components are visible in the in the LOFAR profiles with some evidence of scattering.  So we model the pulsar with a core-cone \textbf{T} beam structure. The fluctuation spectra are featureless.  The BKK19 60-MHz profile is probably highly scattered.  
\vskip 0.1in

\noindent\textit{\textbf{B1842+14}}: The core of this pulsar is not clearly seen at high frequency, but its leading edge seems to be revealed in the LOFAR band. Two conal features are seen at 1.4 GHz and above, but the leading one is barely detectable at 327 MHz and not at all in the LOFAR profiles---so the narrow profiles are problematic. This is probably responsible for the apparent conal narrowing with wavelength.  Taking the core width as twice the leading half-width results in reasonable values down to 129 MHz.  The BKK19 profiles show the onset of scattering---as does the PHS16 38-MHz detection---and only the 71-MHz may be mostly intrinsic.  We modeled this profile as conal because the profile shape seems stable, but it has little intrinsic significance.  Neither Weltevrede \etal\ (\cite{weltevrede2006,weltevrede2007}) nor ourselves find any fluctuation-spectral features.
\vskip 0.1in

\noindent\textit{\textbf{B1845--01}}: This pulsar has a triple profile, so could be either a core-cone triple or a conal triple.  The inner component has too large a width to be a core, and a broad drift feature is detected by Weltevrede \etal\ (\citeyear{weltevrede2006, weltevrede2007}), so we continue to model it as a conal triple.  The scattering is severe in a number of profiles even at 600-MHz (\eg GL98), substantially more so than indicated by the measured level \citep{kmn+15}. However, both the OMR19 327-MHz and MM10 111-MHz profiles show structures that can be interpreted as intrinsic.  Can it be that the scattering in this pulsar is highly time-variable?
\vskip 0.1in

\noindent\textit{\textbf{B1848+12}}:  The OMR19 profiles show a clear triple structure with a very well defined steep PPA traverse.  Moreover the core width seems to be nearly as narrow as the polarcap, but cannot be measured accurately.  This indicates an unusually narrow conal width, and it may be a strong example of a "more inner" cone.  At LOFAR frequencies, the core component is bright, but appears visibly scattered at the lowest frequencies.  See OMR19 for HF profiles.
\vskip 0.1in

\noindent\textit{\textbf{B1848+13}}:  The core component of this \textbf{S$_t$} source increases in width substantially with wavelength. t$_{\rm scatt}$ at LOFAR frequencies may be 3-4\degr and thus account for the width escalation. See OMR19 for HF profiles.
\vskip 0.1in

\noindent\textit{\textbf{B1910+20}}: The \textbf{M}-component structure is discernible only at higher frequencies; see OMR19. The 149-MHz core-width estimate entails a large error.  We confirm Weltevrede \etal's fluctuation spectral feature in the first component and see a weaker one in the trailing component \citep{weltevrede2006, weltevrede2007}.  The central component has a very different behavior and is therefore well identified as a core feature.
\vskip 0.1in

\noindent\textit{\textbf{B1914+09}}; Evan with the broad frequency information we have available, it is difficult to be sure about this pulsar's classification.  We are doubtful that the ET VI model \textbf{$S_{t}$} configuration is correct. A weak leading feature is seen in some profiles that might be conal, which could support the ET VI model.  However, a better model we believe is a conal double, one wherein most core emission is missed by the non-central sightline traverse.  More study is needed.  The LOFAR profile seems to have a scattering tail, compatible with the measured $t_{\rm scatt}$.  
\vskip 0.1in

\noindent\textit{\textbf{B1915+13}}: This pulsar is a well-known \textbf{S$_t$} pulsar, and scattering seems to account fully for the dramatic width escalation below 200 MHz.  Its core and conal widths are difficult to measure accurately, so estimates must suffice.
\vskip 0.1in

\begin{figure}
\begin{center}
\includegraphics[width=80mm,angle=0.]{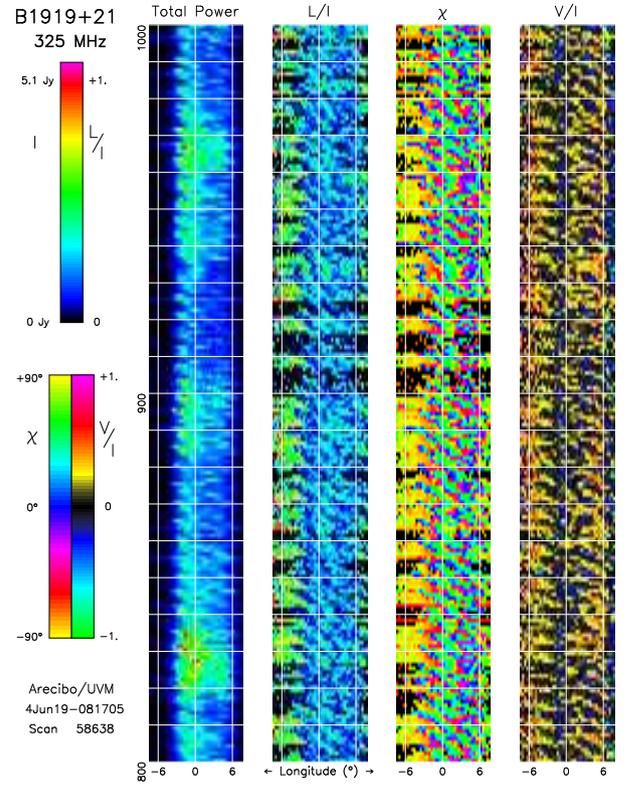}
\caption{A 200-pulse sequence of B1919+21's radio emission---Jocelyn Bell's first pulsar. The total power $I$, fractional linear $L/I$ (top left and right color bars), PPA $\chi$, and fractional circular polarization $V/I$ (bottom left and right color bars) are shown in the four columns according to their respective colour-coded scales on the diagram's left side.  The 3-$\sigma$ background noise level of this sequence is indicated by the white stripe within the lowest intensity black portion of the $I$ color scale, but disappears here due to the large S/N. Notice the prominent subpulse drift in the polarization while the total power has minimal modulation.  Overall, the fractional polarization is low, so most of the total power is unpolarized.  Finally, note the narrow longitude extent of the total power relative to the polarized power.}
\label{figA109}
\end{center}
\end{figure}
\begin{figure}
\begin{center}
\includegraphics[width=75mm,angle=0.]{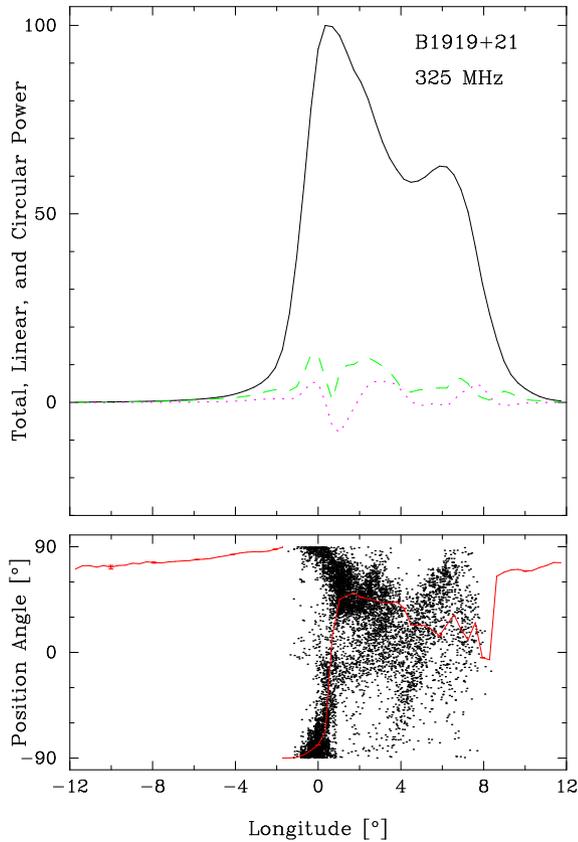}
\caption{B1919+21: Average polarization profile showing the extent of the weak leading linear polarization that defines the PPA far in advance of the profile proper.  Here we see no edge depolarization.  Note also that the PPA rates on the far wings of the profile are positive and shallow by contrast with the usually quoted negative rate associated with the main power under the profile.  Finally, we see little indication of PPAs corresponsing to two OPMs here, though the previous colour polarization shows many red and green orthogonal angles.}    
\label{figA109a}
\end{center}
\end{figure}
\vskip 0.1in
\noindent\textit{\textbf{B1919+21}}: Jocelyn Bell's first pulsar remains exceptional.  It has prominent drifting subpulses, seen mostly in the polarization in Fig.~\ref{figA109}.  Organized subpulse motion extends to much wider longitudes that the total power at 327 MHz. However, there is evidence for two cones in the 1-GHz region, and the \citet{Noutsos} beautifully measured 150-MHz polarized profile may show the inner conal dimension in the linear profile. As seen here, what we have called the outer cone exhibits virtually no increase to the very lowest radio frequencies \citep{mitra2002}---and this ignores the broad extent of organized subpulses.  Also, the PPA sweep rate, and thus the geometry, of this pulsar is difficult to discern due to the small linear polarization under the profile and the change in sweep rate near the wing regions as shown in Fig.~\ref{figA109a}.  We use the \citet{von_thesis} 4.85-GHz profile and the decametric profiles of \citet{Zakharenko2013} without correcting for the 22 ms of scattering at 25 MHz.
\vskip 0.1in

\noindent\textit{\textbf{B1923+04}}: \citet{weltevrede2007} confirms \citet{backus_thesis}  finding a nearly odd-even drift modulation confirming the conal character of the emission in this pulsar.  We therefore model the beams using an inner conal \textbf{S$_d$} geometry.  
\vskip 0.1in

\noindent\textit{\textbf{B1929+10}}: In addition to its prominent core and outer conal components at high frequency, an inner cone can be detected \citet{rankin1997} in this pulsar.  In the decameter band, the core width escalates, possibly due to conflation with the leading conal components.  The pulsar appears to be a two-pole interpulsar with a nearly orthogonal geometry as discussed in OMR19, however the PPA rate is not steep enough to confirm this.  The --6\degr/\degr\ value is then not measured but required to model the beams in the two-pole interpulsar configuration.  We include the decametric profile widths from \citet{Zakharenko2013} as they seem to continue the trend of other low frequency measurements.  It seems that scattering is difficult to measure in this pulsar, as the latter paper gives no value.  We use the \cite{malov} value with some caution.
\vskip 0.1in

\noindent\textit{\textbf{B1933+16}}: This is the brightest core-dominated pulsar in the Arecibo sky, but it can only be studied at high frequency due to scattering.  We confirm the results of the polarimetric, single-pulse study of \cite{mitra2016} in our beam modeling here, correcting the model in OMR19.  The pulsar's complex profile has a core of two parts with different polarization-modal power.  The conal outriders are indistinct at 1.4 GHz with what may be a combination of inner and outer conal power, but at 4.5 GHz they are mostly inner conal.
\vskip 0.1in

\noindent\textit{\textbf{B1944+17}}:  In this pulsar's unusual geometry \citep{kloumann} wherein the sightline circle lies within the conal beam system, the outer cone is seen only at the highest frequency because RFM  increases its radius with wavelength.  A very weak core beam is encountered as the interpulse with a width of some 45\degr, but is seen only at high frequency.  We include the 25-MHz detection by \citep{Zakharenko2013}  without attempting to correct for the 20 ms scattering time.
\vskip 0.1in

\noindent\textit{\textbf{B1935+25}}: As seen in \citet{basu2015} and more recently OMR19, the profile evolves in an unusual manner with 3 components; however, we interpret it as double conal apart from a possible weak core at 327 MHz.  As only outer conal dimensions can be measured, we model it as such.  Were a core beam encountered, it would have a width of about 6\degr.
\vskip 0.1in

\noindent\textit{\textbf{B1946+35}}: This is a well-studied core single \textbf{S$_t$} pulsar with a prominent scattering tail at 300 MHz \citep{Krishnakumar2015}.  The LOFAR 178-MHz shows a nearly half-period scattering tail; however, the MM10 111-MHz profile is much narrower, perhaps due to being ``scattered out''.  The pulsar also shows a non-drift cycle modulating the entire profile \citep{mitra2017}.
\vskip 0.1in

\begin{figure}
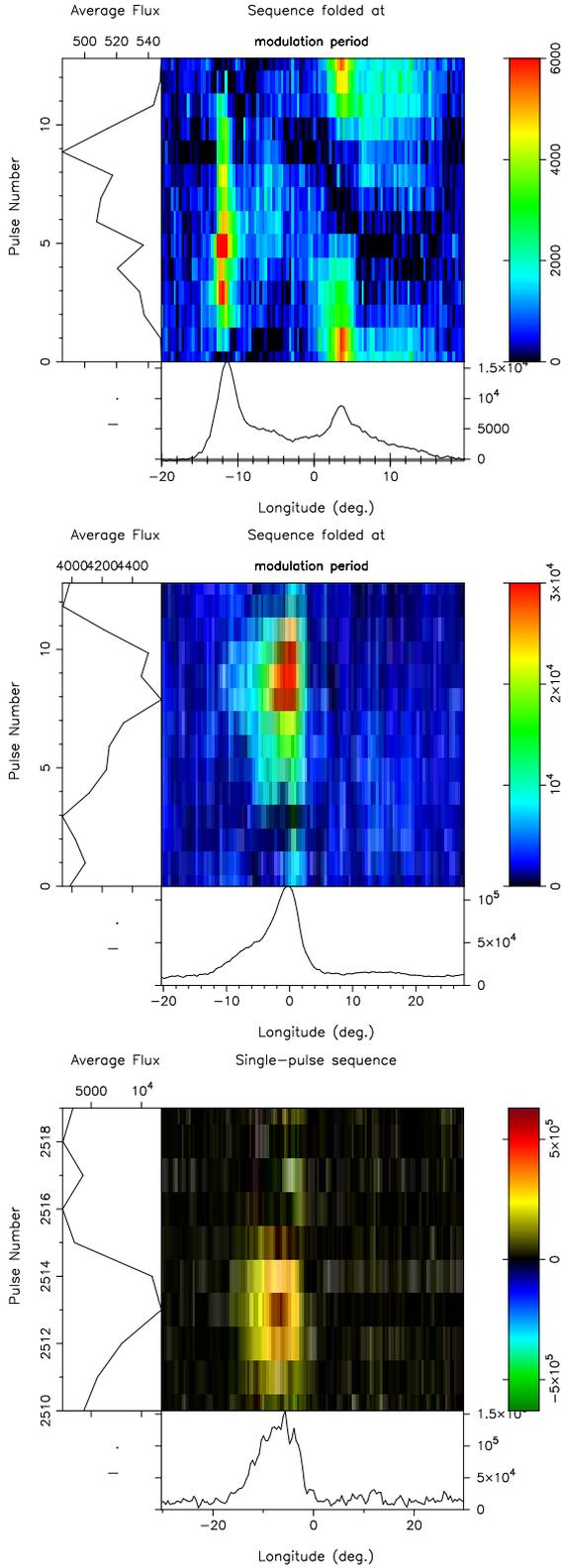

\begin{center}
\includegraphics[width=68mm,angle=-90.]{plots/PQB1952+29.55632la_mod_12.8P.ps}
\vskip 0.05in
\includegraphics[width=68mm,angle=-90.]{plots/PQB1952+29.52947ap_mod_12.8P.ps}
\vskip 0.05in
\includegraphics[width=68mm,angle=-90.]{plots/PQB1952+29.52947ap_giant2513.ps}
\caption{B1952+29:  The top and middle panels show the pulse sequences folded at the 12.8-$P$ cycle at 1.4 GHz and 327 MHz.  Note that different areas of the two profiles are modulated very differently at the two frequencies, but even the weakest parts appear to participate.  Here only the fluctuating power above the stable ``base'' is plotted.  The bottom panel shows a ``giant'' pulse (\#2513) in the 327-MHz observation.  Its intensity is about 60 times that of the typical emission.}
\label{figA110}
\end{center}
\end{figure}

\noindent\textit{\textbf{B1952+29}}: This pulsar has a broad complex profile with very different forms at different frequencies.  The entire profile is modulated on a 12.8-$P$ cycle with different areas illuminated at different phases---as shown in Fig.~\ref{figA110}---so the emission all appears to be conal. The structure seems to reflect a double cone, and no core is clearly discernible in any of our profiles, so we model it with a conal quadruple c\textbf{Q} beam structure.  Were a core beam encountered, it would have a width of about 5.5\degr.
\vskip 0.1in

\noindent\textit{\textbf{B2016+28}}: Prominent drifting subpulses are prominent within this pulsar (\eg \citet{Ramachandran}) and it has a very usual \textbf{S$_d$} outer conal profile. The pulsar was detected at 25 MHz  \citep{Zakharenko2013} but its profile is dominated by scattering.

\noindent\textit{\textbf{B2020+28}}: This pulsar has a well-studied \textbf{D} outer cone profile (\eg \citet{mitra2002}).  However, its leading component shows a sharp polarization-modal boundary and even-odd modulation is seen only in the relatively depolarized trailing component (e.g. \citet{Cordes}). These suggest more complexity than a conal beam system seems able to readily accommodate.  
\vskip 0.1in

\noindent\textit{\textbf{B2110+27}}:  Apparently an \textbf{S$_d$} inner conal pulsar, and the \cite{weltevrede2006,weltevrede2007} analyses support this with their findings about drifting subpulses.  The profiles all have a single form, apart from the surprising conal double form of the high quality \citet{Zakharenko2013}  profile.  We modeled the geometry with an inner cone because little width increase is seen down to the LOFAR band, but the decametric profiles call this into question. 
\vskip 0.1in

\noindent\textit{\textbf{B2303+30}}:  The pulsar has two modes and prominent drifting subpulses \citep{redman} supporting its conal single \textbf{S$_d$} configuration.  Its conal width is so constant down to low frequencies, that an inner cone might be a possibility.
\vskip 0.1in

\begin{figure}
\begin{center}
\includegraphics[width=75mm,angle=-90.]{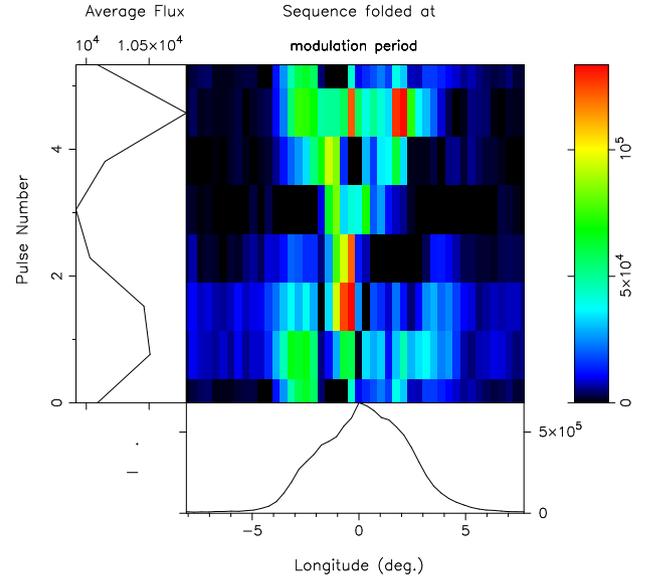}
\caption{B2315+21:  Folded pulse sequence at 327 MHz showing the character of the pulsar's 5.3-$P$ modulation.  The crenellation is not deep because the central component is fairly steady, but both the leading and trailing regions show a strong modulation cycle}
\label{figA112}
\end{center}
\end{figure}
\noindent\textit{\textbf{B2315+21}}: This pulsar shows some evidence of an inner cone at high frequency as well as the prominent outer one, and this squares with the outer cone evolution at low frequency.  We thus model it using a conal triple/quadruple model.  We also find a 5.3-$P$ fluctuation feature at 327 MHz that modulates the entire profile (see Fig.~\ref{figA112}), confirming the result reported by \cite{weltevrede2007}.  The \citet{Zakharenko2013} profile may retain some of its high frequency form despite the scattering.



\bsp	
\label{lastpage}

\end{document}